\pdfoutput=1

\documentclass[11pt,twoside,a4paper,cmspaper,final,collab]{cms-tdr}
\def\svnVersion{435144P}\def\cmsCernNoTag{CERN-PH-EP-2015-189}\def\cmsCernDate{\today}\def\cmsMessage{Published in Physical Review D as \href{http://dx.doi.org/10.1103/PhysRevD.93.034014}{\doi{10.1103/PhysRevD.93.034014.}}}

\begin{document}\cmsNoteHeader{TOP-13-013}

\hyphenation{had-ron-i-za-tion}
\hyphenation{cal-or-i-me-ter}
\hyphenation{de-vices}
\RCS$Revision: 308672 $
\RCS$HeadURL: svn+ssh://svn.cern.ch/reps/tdr2/papers/TOP-13-013/trunk/TOP-13-013.tex $
\RCS$Id: TOP-13-013.tex 308672 2015-11-03 19:45:28Z bbetchar $
\newcolumntype{.}{D{.}{.}{-1}}
\newcolumntype{x}{D{,}{\,\pm\,}{4.21}} 
\newlength\cmsFigWidth
\ifthenelse{\boolean{cms@external}}{\setlength\cmsFigWidth{0.48\textwidth}}{\setlength\cmsFigWidth{0.7\textwidth}}
\ifthenelse{\boolean{cms@external}}{\providecommand{\cmsLeft}{top}\xspace}{\providecommand{\cmsLeft}{left}\xspace}
\ifthenelse{\boolean{cms@external}}{\providecommand{\cmsRight}{bottom}\xspace}{\providecommand{\cmsRight}{right}\xspace}
\newlength\projwidth
\ifthenelse{\boolean{cms@external}}{\setlength\projwidth{0.4\textwidth}}{\setlength\projwidth{0.4\textwidth}}

\renewcommand{\PW}{\ensuremath{\cmsSymbolFace{W}}\xspace}
\renewcommand{\Pe}{\ensuremath{\cmsSymbolFace{e}}\xspace}
\renewcommand{\Pg}{\ensuremath{\cmsSymbolFace{g}}\xspace}
\newcommand{\ejets}{\Pe{}+jets}
\newcommand{\mujets}{\PGm{}+jets}

\newcommand{\QG}{\ensuremath{{\PQq\Pg}}\xspace}
\newcommand{\AG}{\ensuremath{{\PAQq\Pg}}\xspace}
\newcommand{\GG}{\ensuremath{{\Pg\Pg}}\xspace}
\newcommand{\XL}{\ensuremath{\varUpsilon_{\ttbar}}\xspace}
\newcommand{\XLreco}{\ensuremath{\XL^{\text{rec}}}\xspace}

\newcommand{\Eq}[1]{Eq.~(\ref{#1})}
\newcommand{\Eqs}[2]{Eqs.~(\ref{#1},~\ref{#2})}
\newcommand{\Fig}[1]{Fig.~\ref{#1}}
\newcommand{\Tab}[1]{Table~\ref{#1}}
\newcommand{\Ref}[1]{Ref.~\cite{#1}}

\newcommand{\Wj}{\ensuremath{\PW{}\cmsSymbolFace{j}}\xspace}
\newcommand{\DY}{\ensuremath{\mathrm{DY}}\xspace}
\newcommand{\ST}{\ensuremath{\mathrm{St}}\xspace}
\newcommand{\MJ}{\ensuremath{\mathrm{Mj}}\xspace}
\newcommand{\reliso}{\ensuremath{I^\text{rel}_\mathrm{PF}}\xspace}

\newcommand{\KnR}{K{\"u}hn and Rodrigo}
\newcommand{\BnS}{Bernreuther and Si}

\newcommand{\Xmax}{\tilde{X}}
\newcommand{\reco}{\text{rec}}
\newcommand{\MT}{\ensuremath{M_\mathrm{T}}\xspace}
\newcommand{\lepT}{\ensuremath{\ell_\mathrm{T}}\xspace}
\newcommand{\triD}{\ensuremath{\Delta}\xspace}

\cmsNoteHeader{TOP-13-013}
\title{Measurement of the charge asymmetry in top quark pair production in \texorpdfstring{$\Pp\Pp$}{pp} collisions at \texorpdfstring{$\sqrt{s} = 8\TeV$}{sqrt(s)=8 TeV} using a template method}

\date{\today}

\abstract{
The charge asymmetry in the production of top quark and antiquark
pairs is measured in proton-proton collisions at a center-of-mass
energy of $8\TeV$.
The data, corresponding to an integrated luminosity of $19.6\fbinv$,
were collected by the CMS experiment at the LHC.
Events with a single isolated electron or muon, and four or more
jets, at least one of which is likely to have originated from
hadronization of a bottom quark, are selected.
A template technique is used to measure the asymmetry in the
distribution of differences in the top quark and antiquark absolute
rapidities.
The measured asymmetry is $A_c^y = [0.33\pm0.26\stat\pm0.33\syst]\%$, which is the most precise result to date.
The results are compared to calculations based on the standard model and on several
beyond-the-standard-model scenarios.}

\hypersetup{%
pdfauthor={CMS Collaboration},%
pdftitle={Measurement of the charge asymmetry in top quark pair production in pp collisions at sqrt(s) = 8 TeV using a template method},%
pdfsubject={CMS},%
pdfkeywords={top, charge, asymmetry}}

\maketitle

\section{Introduction}

The top quark is the heaviest particle in the standard model
(SM) and the only fermion with a mass on the order of the
electroweak scale~\cite{Agashe:2014kda}.
Deviation of its production or decay properties from the SM
predictions could signal physics beyond the SM.
Several proposed extensions of the SM include heavy mediators of the
strong interaction with axial coupling to quarks, collectively
referred to as axigluons~\cite{Frampton:1987dn}.
Top quark pair production in axigluon-mediated quark-antiquark
annihilation can exhibit a forward-backward asymmetry that depends on
the invariant mass of the system, similar to the asymmetry in fermion
pair production mediated by $\Z$ bosons~\cite{Budny:1973dj}.
These types of models have been leading candidates for accommodating
the behavior of $\ttbar$ production in proton-antiproton collisions
observed by FNAL Tevatron experiments based on about half of their full data set
($5\fbinv$)~\cite{Abazov:2011rq,Aaltonen:2011kc}.
Since analyses of the full Tevatron data set ($10\fbinv$) indicate smaller
values of asymmetry~\cite{PhysRevD.87.092002,Abazov:2014cca},
and since recently improved SM-based theoretical
calculations~\cite{Kuhn:2011ri,PhysRevD.86.034026} predict higher
values of the asymmetry than previous calculations,
the discrepancy between the SM and experimental observations has been
greatly reduced.
Measurements of dijet
production~\cite{Chatrchyan:2013qha,Aad:2014aqa,ATLAS:2012pu} have
constrained the range of axigluon masses and
couplings~\cite{benchmarks},
but the constraints are not applicable to models in which
axigluon-produced dijet resonances are much broader than the
experimental resolution, or which include multiparticle final
states~\cite{Atre:2013mja}.
Precise measurement of the charge asymmetry in top quark pair
production remains one of the best ways to test the limits of validity
of SM predictions.

Experiments at the CERN LHC have reported values of charge asymmetry
in top quark pair
production~\cite{Chatrchyan201228,Chatrchyan2012129,ATLASTTBAR,Aad:2013cea,unfold8TeV}
consistent with SM predictions~\cite{Kuhn:2011ri,PhysRevD.86.034026}.
Corroboration of results from experiments at the Tevatron using
measurements at the LHC is complicated by several differences between
the two colliders.
First, while at the Tevatron the majority of the $\ttbar$ events are
produced via quark-antiquark annihilation, at the LHC the $\ttbar$
production is dominated by charge-symmetric gluon fusion,
$\GG\to\ttbar$.
Second, collisions at the LHC are forward-backward symmetric, so
observation of a charge asymmetry in \ttbar production via
annihilation of a valence quark and a sea antiquark, $\qqbar\to\ttbar$,
relies on the statistical expectation that the system be boosted in
the direction of the quark momentum.
Any difference in top quark and antiquark affinity for the initial
quark or antiquark momentum will consequently result in more forward
production of one and more central production of the other.
This forward-central $\ttbar$ charge asymmetry at the LHC is diluted
relative to the forward-backward $\ttbar$ charge asymmetry at the
Tevatron since the LHC colliding system does not always have a boost
in the expected direction.
Third, a significant portion of LHC $\ttbar$ events are due to
(anti)quark-gluon initial states, $\QG$ ($\AG$), which are charge
asymmetric in number density as well as momentum, and which also
contribute to the final-state forward-central $\ttbar$ asymmetry.
Despite these complications, the large number of $\ttbar$ events
produced at the LHC makes measurement of charge asymmetry competitive
with the Tevatron measurements as a test of the SM.

The measurement of \ttbar asymmetry presented in this paper utilizes a
template technique based on a parametrization of the SM.
The technique differs from previous \ttbar asymmetry
measurements~\cite{Abazov:2011rq,Aaltonen:2011kc,PhysRevD.87.092002,Abazov:2014cca,Chatrchyan201228,Chatrchyan2012129,ATLASTTBAR,Aad:2013cea,unfold8TeV},
which are based on unfolding the effects of selection and resolution
in the observable distribution.
Reference \cite{unfold8TeV} in particular analyzes the same data set,
but also differs in selecting fewer events with higher purity as a
result of more restrictive jet transverse momentum criteria, and in the methods used
to reconstruct \ttbar kinematics and determine the sample
composition.

The template technique is presented in Section \ref{strategy_section}.
Data from
proton-proton collisions at $\sqrt{s}=8\TeV$ were collected in 2012 by the
CMS experiment,  described in Section \ref{cms_section}.
Event selection, reconstruction of \ttbar kinematics, and a population
discriminant are described in Section \ref{selection_section}.
The details of the model used to obtain the result are given in
Section \ref{measurement_section}, and the result is presented in
Section \ref{results_section}.
The analysis is summarized in Section \ref{summary}.

\section{Analysis strategy}
\label{strategy_section}
Charge asymmetry in \ttbar production can be defined for an observable
$X$ that changes sign under the exchange $\PQt\leftrightarrow\PAQt$.
If $X$ is distributed with a differential cross section $\rd\sigma/\rd X$, its probability
density is
\begin{linenomath}
\begin{equation}
\rho(X) = \frac{1}{\sigma}\frac{\rd\sigma}{\rd X}.
\end{equation}
\end{linenomath}
This can be expressed as a sum of symmetric ($\rho^+$) and
antisymmetric ($\rho^-$) components,
\begin{linenomath}
\begin{equation}
  \label{rho_pm}
  \rho^\pm(X) = \left[\rho(X) \pm \rho(-X)\right] / 2.
\end{equation}
\end{linenomath}
Statistical kinematic differences between top quarks and antiquarks
can be summarized in a charge asymmetry,
\begin{linenomath}
\begin{equation}
\label{charge_asymmetry}
A_c^X = \int_0^{\Xmax}\rho(X)\,\rd X - \int_{-\Xmax}^0\rho(X)\,\rd X = 2\int_0^{\Xmax}\rho^-(X)\,\rd X,
\end{equation}
\end{linenomath}
where the observable's maximum value $\Xmax$ may be finite or infinite.
Previous LHC analyses~\cite{Chatrchyan201228,Chatrchyan2012129,ATLASTTBAR,Aad:2013cea,unfold8TeV}
defined a \ttbar charge asymmetry $A_c^y$, based on the difference in
absolute rapidities of the top quark ($y_{\PQt}$) and antiquark
($y_{\PAQt}$), \newcommand{\deltaAbsY}{\Delta\abs{y}_{\ttbar}}
\begin{linenomath}
\begin{equation}
  \label{deltaAbsY}
  \deltaAbsY = \abs{y_{\PQt}} - \abs{y_{\PAQt}}.
\end{equation}
\end{linenomath}
For the technique described in this paper, it is
desirable that the observable $X$ be bounded.
The hyperbolic tangent is a symmetric and monotonic function, so the
transformed observable
\begin{linenomath}
\begin{equation}
\label{XL}
\XL = \tanh\deltaAbsY,
\end{equation}
\end{linenomath}
has the asymmetry $A_c^y$ and is also bounded.

Charge asymmetries at production can only be determined from observed
data distributions using an extrapolation based on a particular model.
Past measurements were extrapolated using an unfolding technique, which
relies on a model for the selection efficiencies and reconstruction
effects~\cite{Abazov:2011rq,Aaltonen:2011kc,PhysRevD.87.092002,Abazov:2014cca,Chatrchyan201228,Chatrchyan2012129,ATLASTTBAR,Aad:2013cea,unfold8TeV}.
An alternative extrapolation discussed in this paper uses a model to
derive template distributions for the symmetric and antisymmetric
components, $\rho^\pm$.

In the present analysis, the next-to-leading-order (NLO) \POWHEG event
generator (version 1.0)~\cite{Frixione:2007nw} is used in association
with the CT10~\cite{Lai:2010vv} parton distribution functions (PDFs) as
a base model to construct the symmetric and antisymmetric components
of the probability density $\rho(X)$ for an observable $X$.
These distributions are represented as symmetrically binned histograms,
given as vectors $\vec{x}^\pm$ with a dimensionality equal to
the number of bins.
A generalized model with a single parameter $\alpha$ can be
constructed from a linear combination of the base model components,
\begin{linenomath}
\begin{equation}
  \label{alpha_model}
\vec{x}^\alpha = \vec{x}^+ + \alpha\vec{x}^-.
\end{equation}
\end{linenomath}
The measurement strategy is to find the value of $\alpha$ that best
fits the observations.
The base model charge asymmetry $\hat{A}_c^X$ is given by
\Eq{charge_asymmetry}.
The charge asymmetry observed in data is then equal to that of the
base model scaled by the parameter $\alpha$:
\begin{linenomath}
\begin{equation}
  \label{extrapolated_asymmetry}
  A_c^X(\alpha) = \alpha\hat{A}_c^X.
\end{equation}
\end{linenomath}

\begin{table}[htb]
  \centering
  \topcaption{\label{powheg_asymmetries}
The \ttbar initial-state fractions and charge asymmetries in the
observable $\XL$, calculated with \POWHEG using the CT10
PDFs.
The statistical uncertainty  in the last digits is indicated in parentheses.
}
  \begin{scotch}{c..}
    Initial state & \multicolumn{1}{c}{Fraction\,(\%)}  & \multicolumn{1}{r}{$\hat{A}_c^y$\,(\%)} \\
    \hline
    $\Pg\Pg$  &   65.2  &-0.059(25) \\
    $\Pq\Paq$ &   13.4   &   2.95(6) \\
    $\Pq\Pg$  &   18.2   &   1.17(5) \\
    $\Paq\Pg$ &    3.2   &-0.21(11)  \\
    \hline
    $\Pp\Pp$  &   100.0       &   0.563(20) \\
  \end{scotch}
\end{table}

\begin{figure}[htb]
  \centering
  \includegraphics[width=0.48\textwidth]{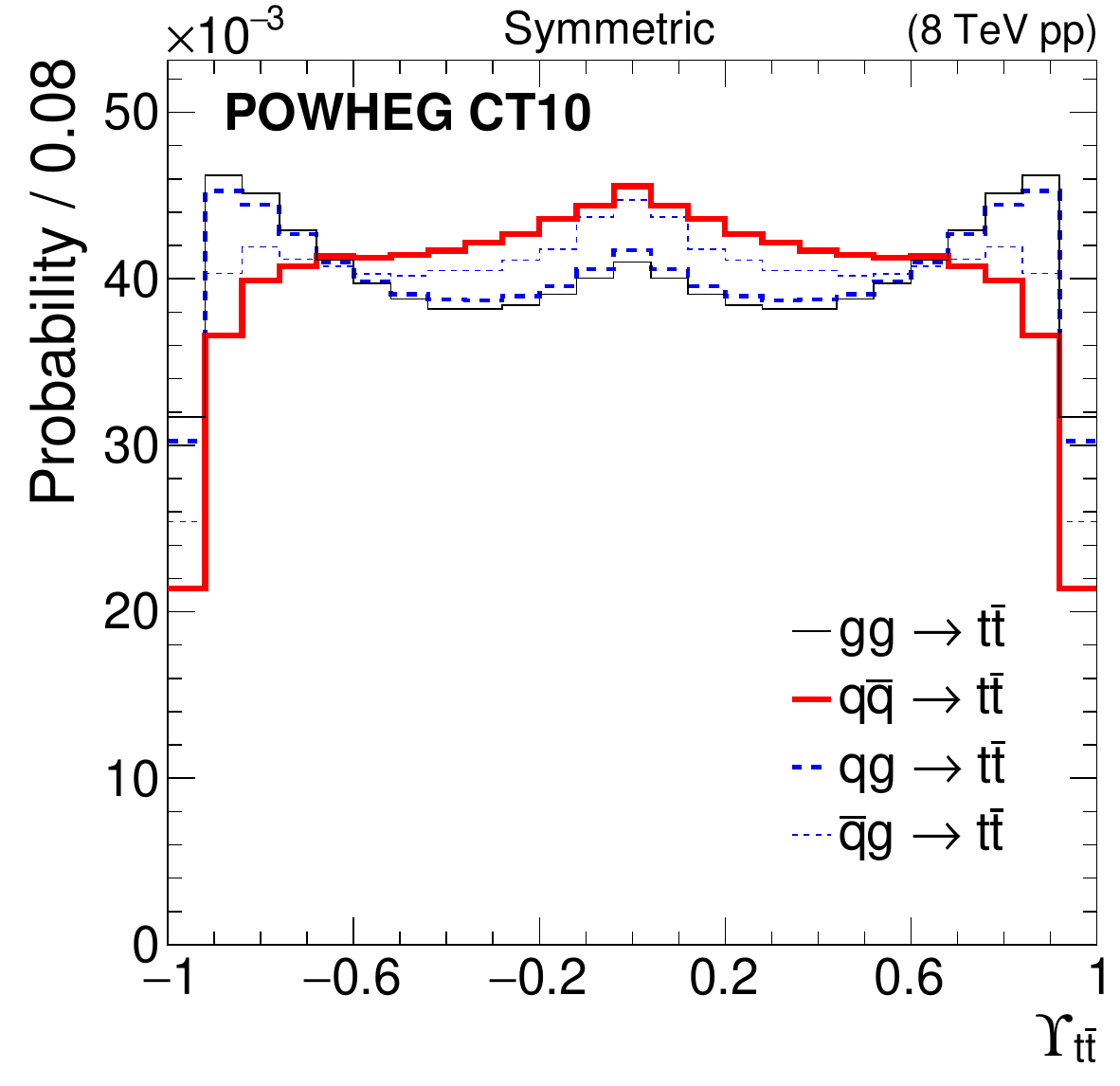}
  \includegraphics[width=0.48\textwidth]{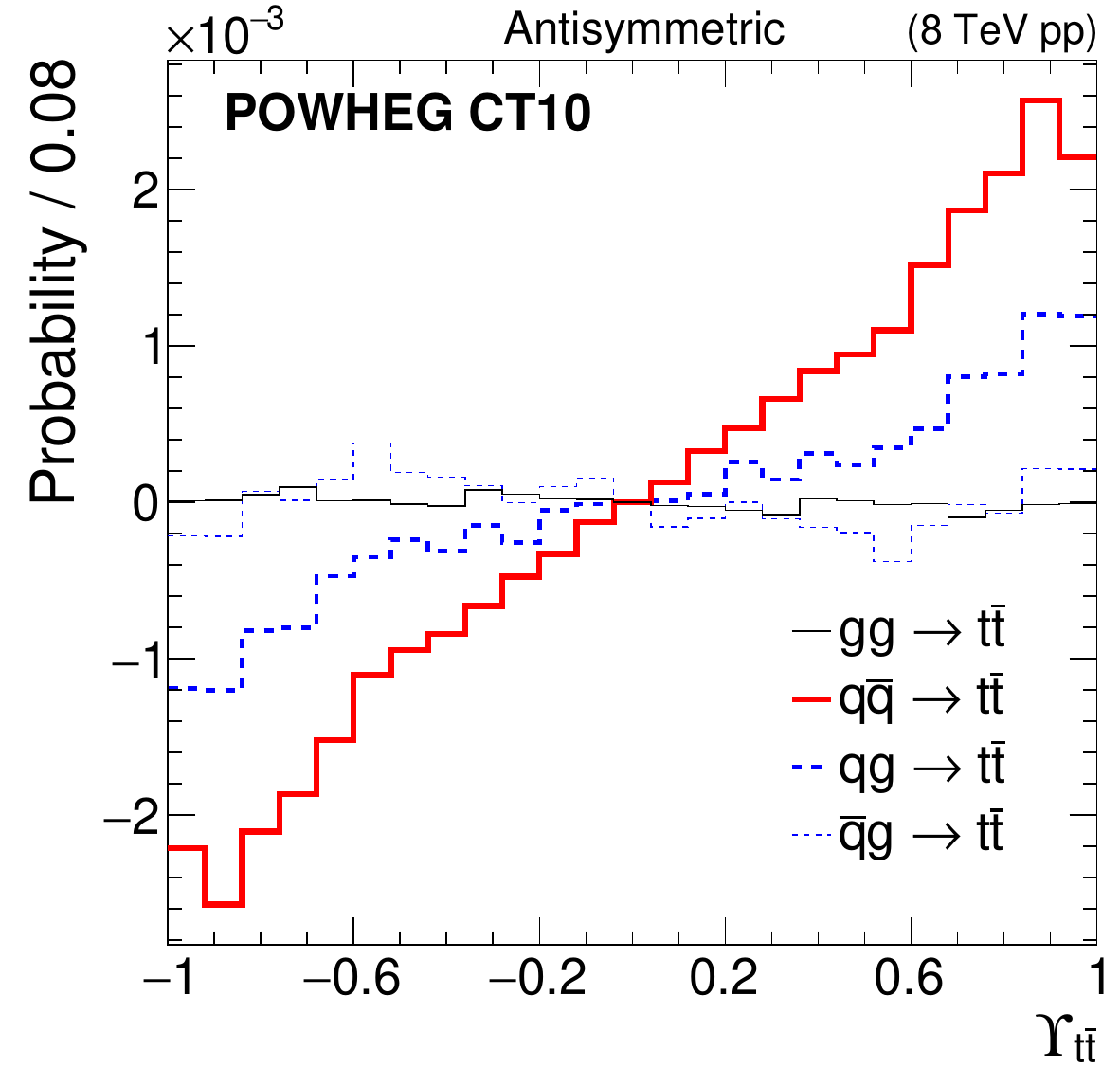}
\caption{\label{Xsymmanti}
The (\cmsLeft) symmetric $\vec{x}^+$ and (\cmsRight) antisymmetric $\vec{x}^-$ components of
the binned probability distributions in the observable $\XL$,
constructed using \POWHEG~\cite{Frixione:2007nw} with CT10
PDFs~\cite{Lai:2010vv}, for \ttbar production from $\GG$, $\qqbar$, $\QG$,
and $\AG$ initial states.}
\end{figure}

Figure \ref{Xsymmanti} presents the $\vec{x}^\pm$ distributions in
$\GG$, $\qqbar$, and $\QG$ ($\AG$) initial states for $X=\XL$, before the
event reconstruction and selection are applied, and the composition
and intrinsic charge asymmetries of each initial state are listed in
\Tab{powheg_asymmetries}.
Imperfect detector resolution, event reconstruction, and selections
can result in distributions of the reconstructed observable $\XLreco$
that differ from those in $\XL$.
For this reason, the symmetric and antisymmetric templates,
$\vec{x}^\pm_{\text{rec}}$, are constructed using \POWHEG-generated
events that are fully reconstructed and pass the selection
criteria.
Studies of simulated events show that event reconstruction and
selection may amplify or dilute an underlying asymmetry in the
$\XLreco$ distribution but do not introduce a significant false bias.
Thus, the scale parameter $\alpha$ in \Eqs{alpha_model}{extrapolated_asymmetry} can
be determined by a fit to the reconstructed distribution in data,
\begin{linenomath}
\begin{equation}
  \label{amplitude_reduced}
  \vec{x}^\alpha_{\text{data}} = \vec{x}^+_{\text{rec}} +
\alpha\vec{x}^-_{\text{rec}}.
\end{equation}
\end{linenomath}

\section{CMS detector and definition of physics objects}
\label{cms_section}
The central feature of the CMS apparatus is a superconducting solenoid
of 6\unit{m} internal diameter, providing a magnetic field of
3.8\unit{T}. Within the solenoid volume are a silicon
pixel and strip tracker, a lead tungstate crystal electromagnetic
calorimeter (ECAL), and a brass and scintillator hadron calorimeter
(HCAL), each composed of a barrel and two endcap sections. Muons are
measured in gas-ionization detectors embedded in the steel flux-return
yoke outside the solenoid. Extensive forward calorimetry complements
the coverage provided by the barrel and endcap detectors.

The first level of the CMS trigger system, composed of custom
hardware processors, uses information from the calorimeters and muon
detectors to select the most interesting events in a fixed time
interval of less than 4\mus. The high-level trigger processor
farm further decreases the event rate from around 100\unit{kHz} to
around 400\unit{Hz}, before data storage.
Single-electron and single-muon triggers were used to collect events
for this analysis.

The particle-flow event
algorithm~\cite{CMS-PAS-PFT-09-001,CMS-PAS-PFT-10-001} is used to
reconstruct and identify each individual particle with an optimized
combination of information from the various elements of the CMS
detector.
Photons and electrons are defined as clusters in ECAL with a
requirement that there be a charged-particle trajectory pointing to an
electron cluster.
The energy of a photon is directly obtained from the ECAL measurement,
corrected for zero-suppression effects.
The energy of an electron is determined from a combination of the
electron momentum at the primary interaction vertex as determined by
the tracker, the energy of the corresponding ECAL cluster, and the
energy sum of all bremsstrahlung photons spatially compatible with
originating from the electron track~\cite{Khachatryan:2015hwa}.
The momentum of a muon is obtained from the direction and curvature of
its combined trajectory in the muon and tracking systems.
The energy of a charged hadron is determined from a combination of
its momentum measured in the tracker and the matching ECAL and HCAL
energy deposits, corrected for zero-suppression effects and for the
response function of the calorimeters to hadronic showers. Finally,
the energy of a neutral hadron is obtained from the corresponding
corrected ECAL and HCAL energy deposits.

For each event,
after identification and removal of leptons relevant to the sample
selection and particles from additional proton-proton interactions
within the same bunch crossing (pileup),
hadronic jets are clustered from these reconstructed
particles with the infrared- and collinear-safe
anti-\kt algorithm, operated with a size parameter $R$ of
0.5~\cite{Cacciari:2008gp}. The jet momentum is determined as the
vectorial sum of all particle momenta in this jet, and is found in the
simulation to be within 5 to 10\% of the true momentum over the
whole transverse momentum (\pt) spectrum and detector acceptance.
Jet energy corrections are derived from the simulation,
and are confirmed with in situ measurements of the energy balance of
dijet and photon+jet events~\cite{Chatrchyan:2011ds}. The jet energy
resolution amounts typically to 15\% at 10\GeV, 8\% at 100\GeV, and
4\% at 1\TeV.
An offset correction is applied to jet energies to take into account pileup contributions.
Additional selection criteria are applied to each event to remove
spurious jet-like features originating from isolated noise patterns in
certain HCAL regions.
Jets from \PQb quarks are identified using a discriminant
containing information about secondary vertices formed by at least
three charged-particle tracks, including the number of associated
tracks, the displacement from the collision point, and the vertex
mass, which is computed from the tracks associated with the secondary
vertex~\cite{1748-0221-8-04-P04013}.

The missing transverse momentum vector \ptvecmiss is defined as the
projection on the plane perpendicular to the beams of the negative
vector sum of the momenta of all reconstructed particles in an
event. Its magnitude is referred to as \ETmiss.

A more detailed description of the CMS detector, together with a
definition of the coordinate system used and the relevant kinematic
variables, can be found in Ref.~\cite{Chatrchyan:2008zzk}.

\section{Event selection and reconstruction}
\label{selection_section}
Each event is considered under the hypothesis that a top quark and a top
antiquark each decay into a bottom quark and a $\PW$ boson, and
that one $\PW$ boson subsequently decays into a pair of quarks, while the
other decays into a neutrino and either an electron or a muon, producing
a lepton and jets ($\ell$+jets) signature.

Events are selected from data collected from collisions of protons at
$8\TeV$ center-of-mass energy and corresponding to an integrated
luminosity of $(19.6\pm0.5)\fbinv$~\cite{Leonard:2014qda}.
Selected events contain at least four jets each with $\abs{\eta}<2.5$ and
$\pt>20\GeV$, one isolated electron (muon) with $\abs{\eta}<2.5$
($2.1$) and $\pt>30\ (26)\ \GeV$.
Events are also required to have no other electrons
($\abs{\eta}<2.5$, $\pt>20\GeV$) or muons ($\abs{\eta}<2.5$,
$\pt>10\GeV$).
A selected event must have an electron with a particle-flow relative
isolation \reliso less than 0.1, or a muon with \reliso less than
0.12~\cite{Khachatryan:2015hwa, Chatrchyan:2012xi}.
Events containing an electron with $0.11<\reliso<0.15$ or a muon with
$0.13 < \reliso < 0.20$ are retained as a control, or sideband,
region.
The (next-to-) leading jet must have $\pt>45\ (35)\ \GeV$.
At least one jet must be \PQb-tagged, as defined by the medium
working point of the combined secondary vertex \PQb tagging
discriminant (CSV), which has an efficiency better than about 65\% and
a misidentification probability of about
1.5\%~\cite{1748-0221-8-04-P04013}.
In total, 326\,185 events are accepted with an electron and jets in the
final state, hereafter referred to as the \ejets\ channel, and
340\,911 events are accepted in the \mujets\ channel.

In addition to \ttbar production, several other processes can produce a
$\ell$+jets signature that passes this selection.
In particular, these processes include production of leptonically
decaying \PW bosons in association with jets (\Wj), Drell--Yan (\DY) production
of $\ell^+\ell^-$ pairs from \qqbar annihilation in
association with jets and in which one lepton is not identified, and
the production of single top (\ST) quarks accompanied by additional jets.
Production of quantum chromodynamic multijets (\MJ) also contributes to the background.
Such events can satisfy the selection if a jet is misidentified as an
electron or if a muon produced in the decay of a heavy quark passes
the isolation criteria.

More than $65\%$ of selected events contain \ttbar pairs.

\subsection{Modeling of signal and background}
\label{signal_templates}

The detection of generated particles is fully simulated with the
\GEANTfour software~\cite{Agostinelli:2002hh} using a detailed description of
the CMS detector.
The samples account for the observed multiplicity of pileup
interactions in data.
Additional weights are applied after event selection to match
the efficiency of triggers and object identification that are
measured in a data sample of $\Z$+jets events using a tag-and-probe
method~\cite{Khachatryan:2015hwa,Chatrchyan:2012xi}.
The energy difference between each reconstructed jet and its
corresponding generated jet is scaled to match the ($\eta$- and
$\pt$-dependent) jet energy resolution in data, as measured using the
dijet asymmetry technique~\cite{Chatrchyan:2011ds}.

As mentioned, the \ttbar events are generated with the NLO \POWHEG
heavy-quark pair production algorithm, using the CT10 PDFs, and interfaced
with \PYTHIA~(version 6.426) for parton showering and
hadronization~\cite{Sjostrand:2006za,Sjostrand:2007gs,Corke:2010zj}.
Events with $\PW$ or \Z bosons in conjunction with 1, 2, 3, or 4 jets
are generated with leading-order (LO) \MADGRAPH~(version
5.1.3.30)~\cite{Alwall:2011uj}, using the CTEQ6
PDFs~\cite{Pumplin:2005rh} (version L1), and are interfaced
with \PYTHIA.
A dedicated $\PW$+$\bbbar$ sample is used for investigation
of systematic uncertainties.
Events with single top quarks or antiquarks are generated with \POWHEG
using the CTEQ6 PDFs (version M),
in the $s$ and $t$ channels~\cite{Alioli:2009je}, and in
the $\PQt\PW$ channel using diagram removal rather than the diagram
subtraction method~\cite{Re:2010bp}.

The \MJ background has a very low efficiency to pass the
selection, making it difficult to simulate enough selected events, but
it has a large enough cross section to make it significant.
The \MJ background is modeled using the sideband data,
subtracting the contributions of simulated processes, which are
normalized according to the integrated luminosity and their cross
sections and selection efficiencies.

Several alternative models of \ttbar production are used to
investigate systematic uncertainties and to evaluate the performance
of the method.
Alternative SM \ttbar simulations are generated with \MADGRAPH and
with \MCATNLO (version 3.41)~\cite{Frixione:2002ik} using the CTEQ6
PDFs~(versions L1 and M, respectively).
Systematic uncertainties related to the factorization and
renormalization scales are evaluated using \POWHEG \ttbar samples in
which both scales are increased or decreased simultaneously by a
factor of two from their nominal values, equal to the event
momentum transfer squared; these control samples are processed with
the \textsc{FastSim}~\cite{Giammanco:2014bza} simulation of the CMS
detector.
A set of six
models in which \ttbar production kinematics are modified by the presence of new physics
are generated with \MADGRAPH, and are described in detail in
\Ref{benchmarks}.
The models are chosen to have parameters not yet excluded by other
experimental constraints.
The set includes a model with an added complex gauge boson
\PZpr~\cite{Jung:2011zv} with a mass of 220\GeV and a coupling to
right-handed up-type quarks.
Other models in the set include parametrized color-octet vector bosons
(axigluon) models~\cite{Frampton:1987dn}, in which the axigluon has
nonzero mass and chiral couplings.
Three models include a light axigluon with a 200\GeV mass and
coupling characterized as right, left, or axial.
Two models include a heavy axigluon with a 2\TeV mass and right or
axial coupling.

\subsection{Reconstruction of top quarks}
\label{ttreco}

Top quarks are reconstructed using the most likely assignment of the
reconstructed jets to the \ttbar decay partons.
Jet four-momenta are corrected according to their parton assignment
and a kinematic fit, which uses the known top quark and $\PW$ boson
masses~\cite{Agashe:2014kda}.
The neutrino momentum is calculated analytically~\cite{Betchart:2013nba}.
The top quark and antiquark four-momenta are found by summing the four-momenta of
their respective decay products.
The charge of the leptonically decaying top quark is determined by
that of the electron or muon, while the top quark that decays into jets
is assumed to be of the opposite charge.

All jet assignments are considered in selecting the assignment of
maximum likelihood.
The selection ensures that the number of jets in the event $N_j$ is at
least four.
There are $N_c=\frac{1}{2}N_j!/(N_j-4)!$, or a minimum of 12, possible jet assignment combinations.
Each assignment is represented by a tuple $(a,b,c,d,\{x\})$, where $a$
represents the \PQb jet associated with $\PQt\to
\PQb\ell\nu_\ell$ decay, $b$ represents the \PQb jet
associated with $\PQt\to\PQb\qqbar$ decay, $c$ and $d$ represent
the two jets from hadronic $\PW$ boson decay, ordered by \pt, and
$\{x\}$ represents any additional jets in the event, ordered by
\pt.
The correct assignment in simulation is designated
$(\hat{a},\hat{b},\hat{c},\hat{d},\{\hat{x}\})$.

\begin{figure*}[htb]
\centering
\includegraphics[width=0.32\linewidth]{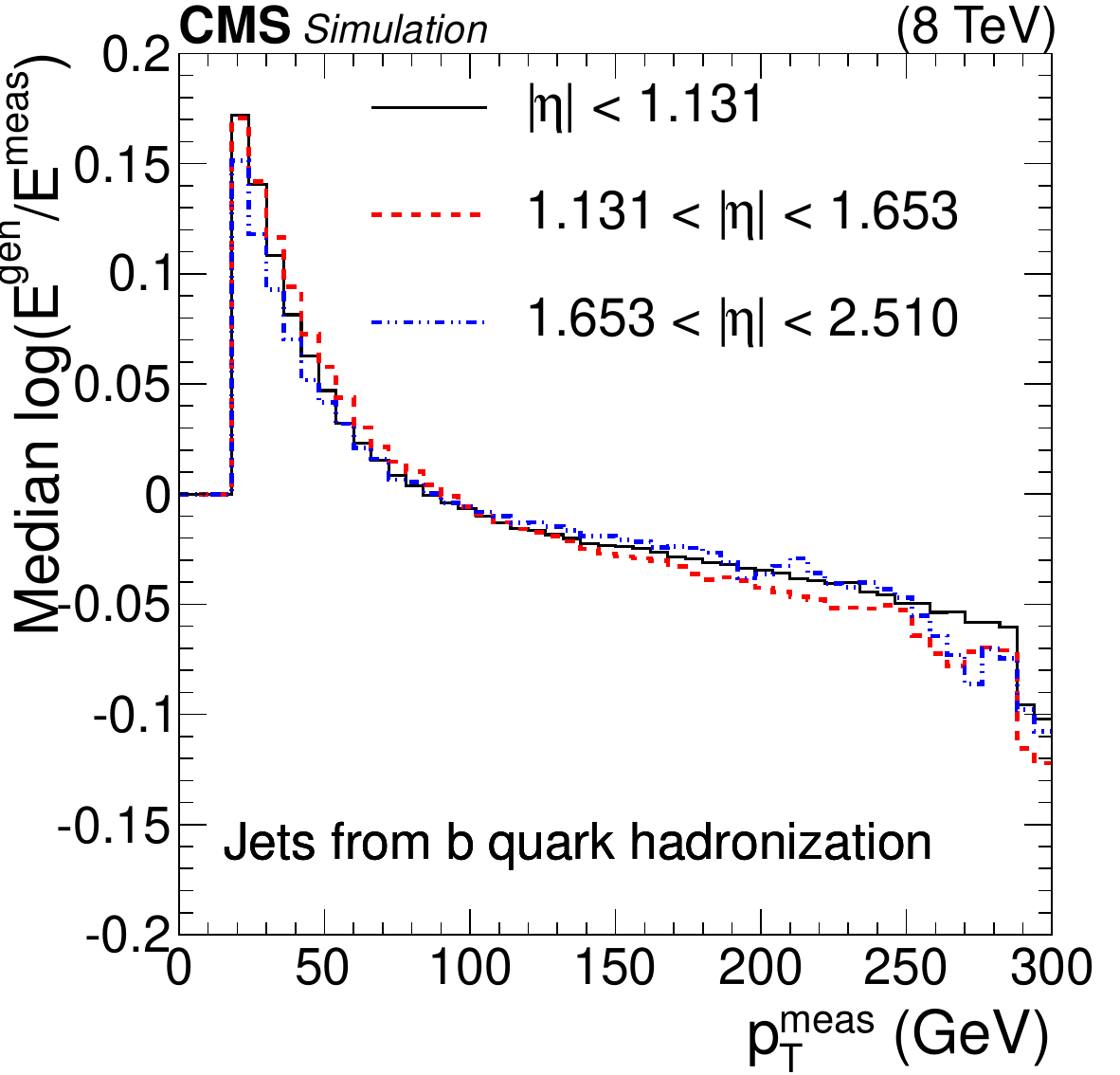}%
\includegraphics[width=0.32\linewidth]{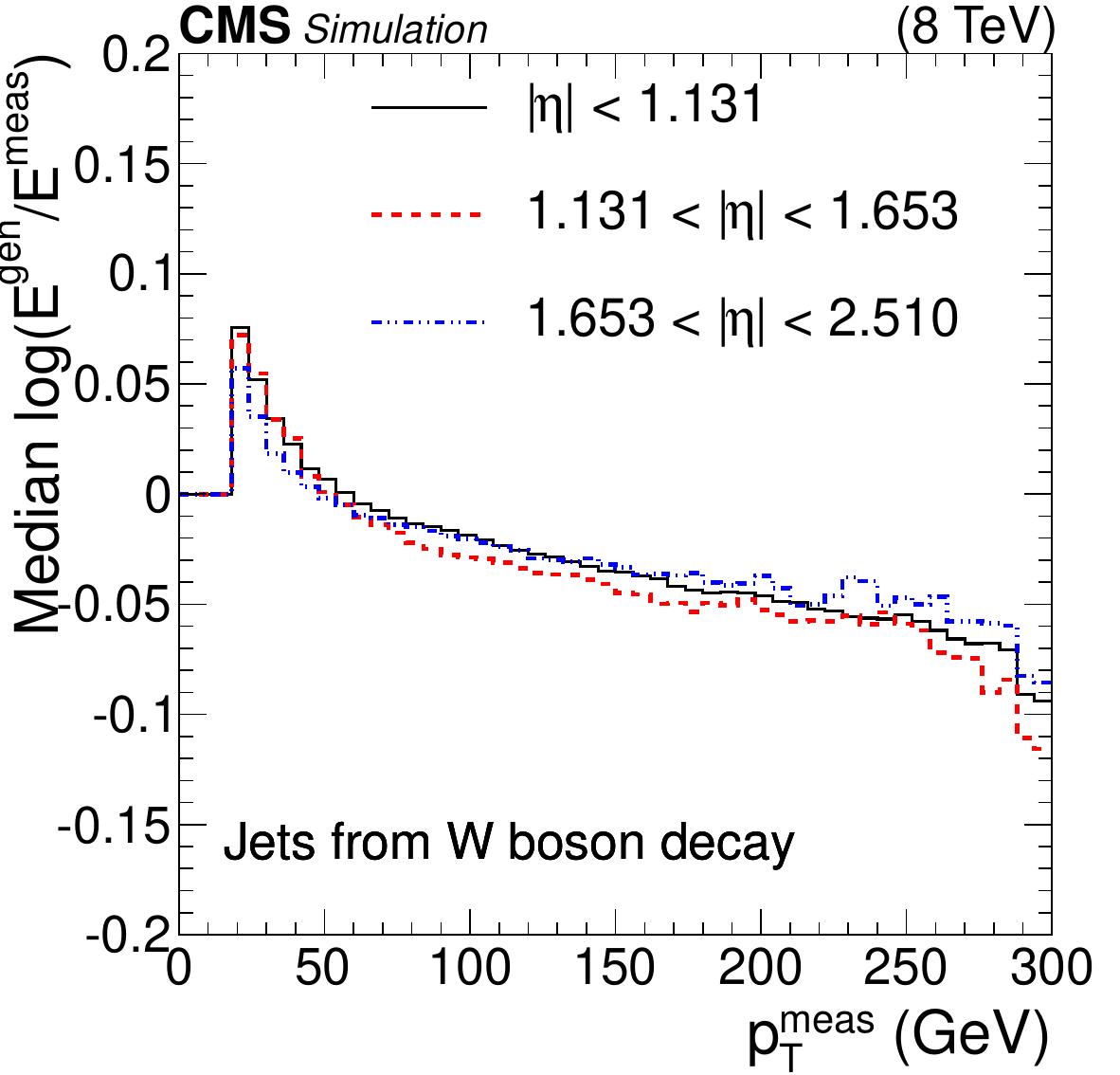}%
\includegraphics[width=0.32\linewidth]{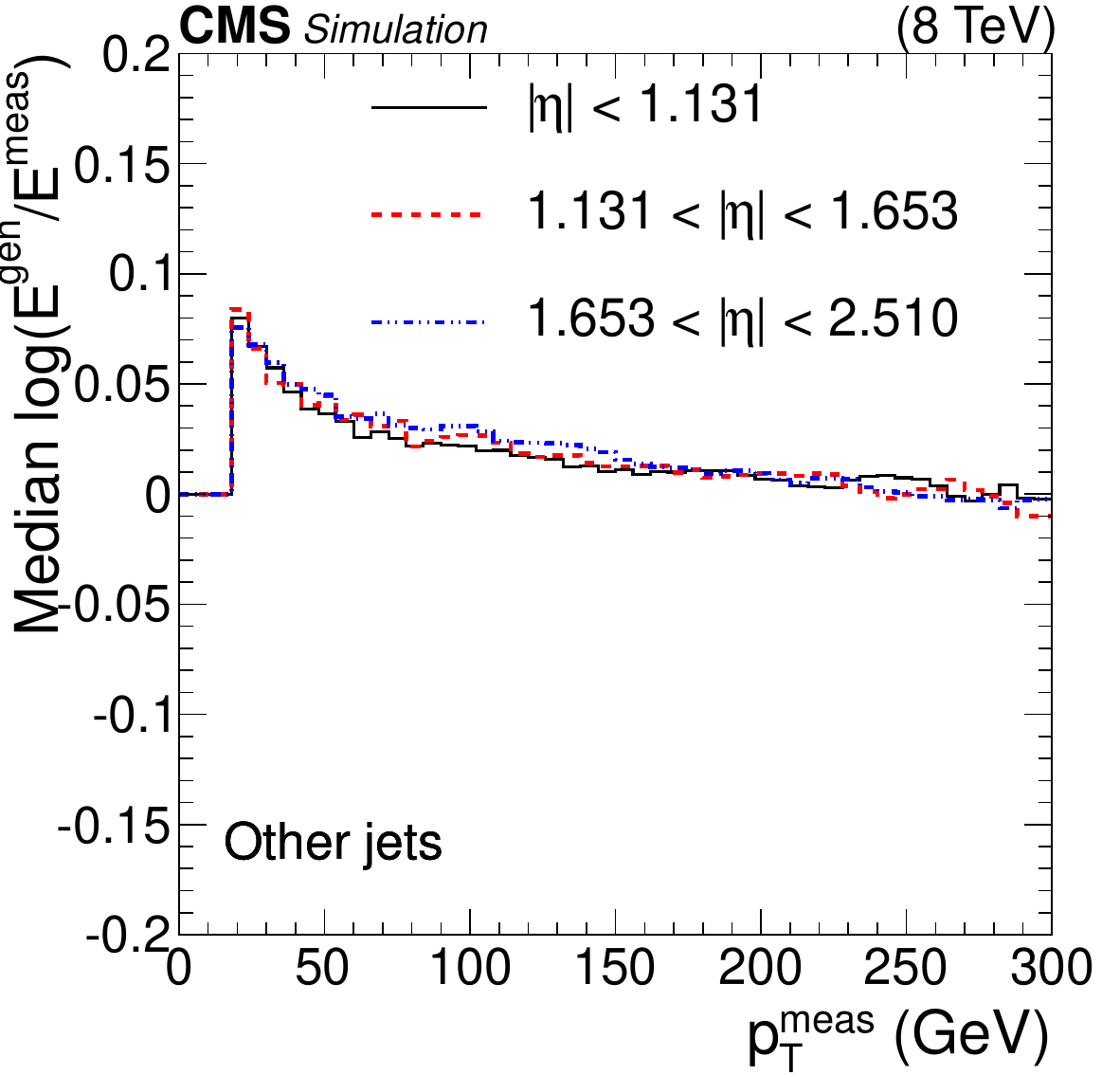}
\caption{\label{jetScaling}
The median value of the logarithm of the ratio of parton energy to
measured energy, as a function of measured \pt in
three bins of $\abs{\eta}$, for (left)
\PQb jets from top quark decay, (center) jets from
$\PW$ boson decay, and (right) other jets.}
\end{figure*}

The scale factors for correcting the energy of the jets from the
reconstruction to the parton level are obtained from $\ttbar$
simulation, following the event selection, for \PQb jets from top quark
decay, jets from $\PW$ boson decay, and other jets.
Corrections are found as a function of \pt in three bins of absolute
pseudorapidity, with upper bin boundaries at $\abs{\eta} = $ 1.131,
1.653, and 2.510, corresponding to the calorimeter barrel, transition,
and endcap regions.
The corrections, shown in \Fig{jetScaling}, are applied to the
measured jet energies according to the assignment.

The likelihood of a given jet-to-parton assignment $i$ is
\begin{linenomath}
\begin{equation}
L_i = L^{\mathrm{CSV}}_i \, LR^{\mathrm{MSD}}_i \,
LR^{\chi}_i,
\end{equation}
\end{linenomath}
where $L^{\mathrm{CSV}}_i$ is the likelihood of the jet \PQb tagging
discriminants, $LR^{\mathrm{MSD}}_i$ is the likelihood ratio of the
invariant masses of jet combinations associated with $\PQt\to\PQb\qqbar$
decays, and $LR^{\chi}_i$ is the likelihood ratio of the $\chi^2$
associated with the products from $\PQt\to\PQb\ell\nu_\ell$ decays.

\begin{figure}[htb]
  \centering
  \includegraphics[width=0.49\textwidth]{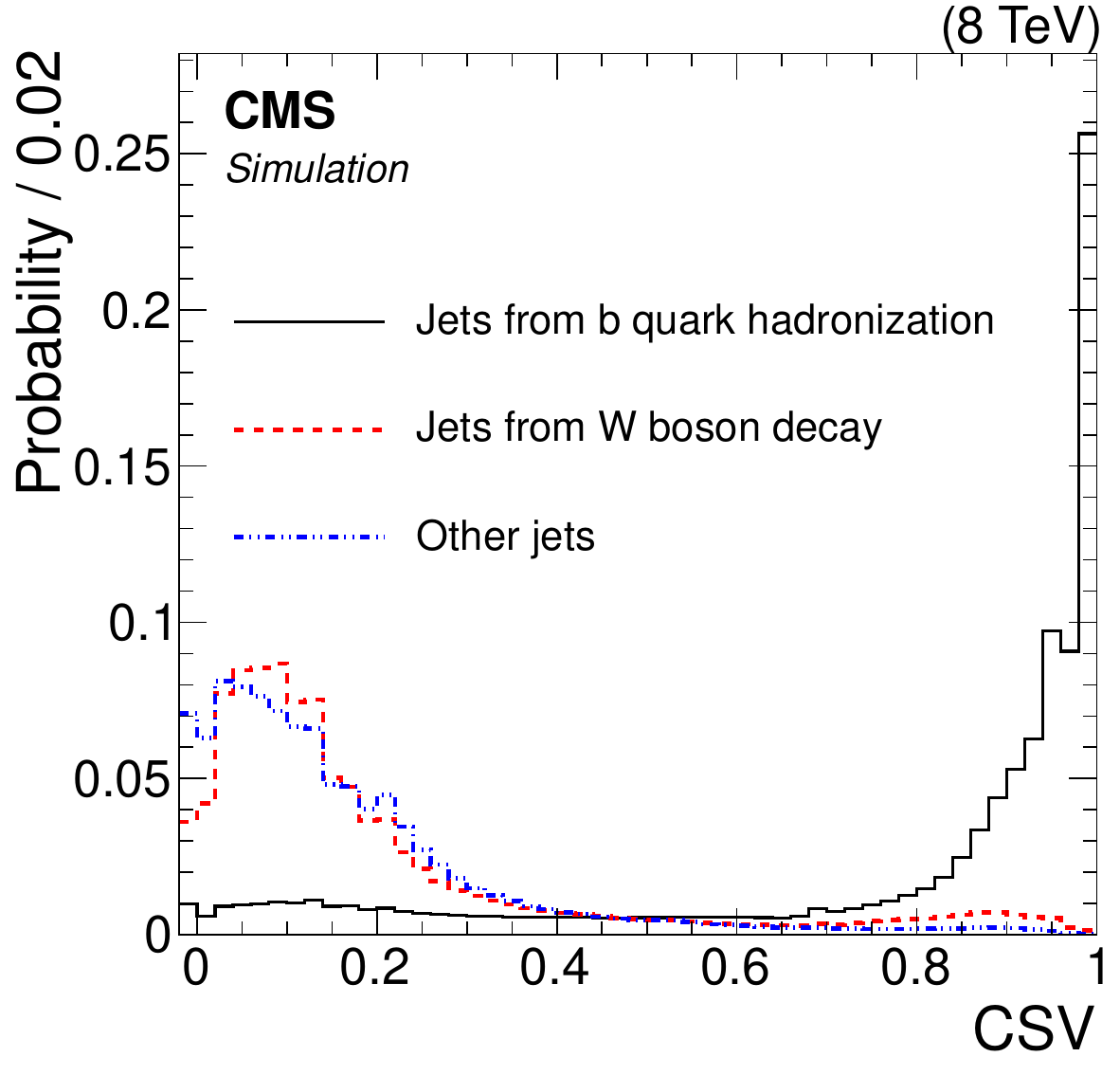}
  \caption{\label{csv_conditional} The conditional probability
    densities of the CSV \PQb tagging discriminant from simulation for jets
    from \PQb quarks, jets from $\PW$ boson decay,
    and other jets.}
\end{figure}

The CSV \PQb tagging discriminant associates a value $\beta$ with
each jet.
The conditional CSV probability densities $\mathcal{B} =
\rho(\beta|\hat{a},\hat{b})$, $\mathcal{Q}=\rho(\beta|
\hat{c},\hat{d})$, and $\mathcal{N}=\rho(\beta|\{\hat{x}\})$ are shown
in \Fig{csv_conditional}.
The likelihood of a given jet assignment $i$, considering the associated
CSV values $\{\beta\}$, is
\begin{linenomath}
\begin{equation}
  \label{lCSV}
  L^{\mathrm{CSV}}_i =
  \mathcal{B}(\beta_a) \mathcal{B}(\beta_b) \mathcal{Q}(\beta_c) \mathcal{Q}(\beta_d) \prod_{j\in\{x\}}\mathcal{N}(\beta_{j}).
\end{equation}
\end{linenomath}

\begin{figure*}[htbp]
  \centering
  \includegraphics[width=0.4\linewidth]{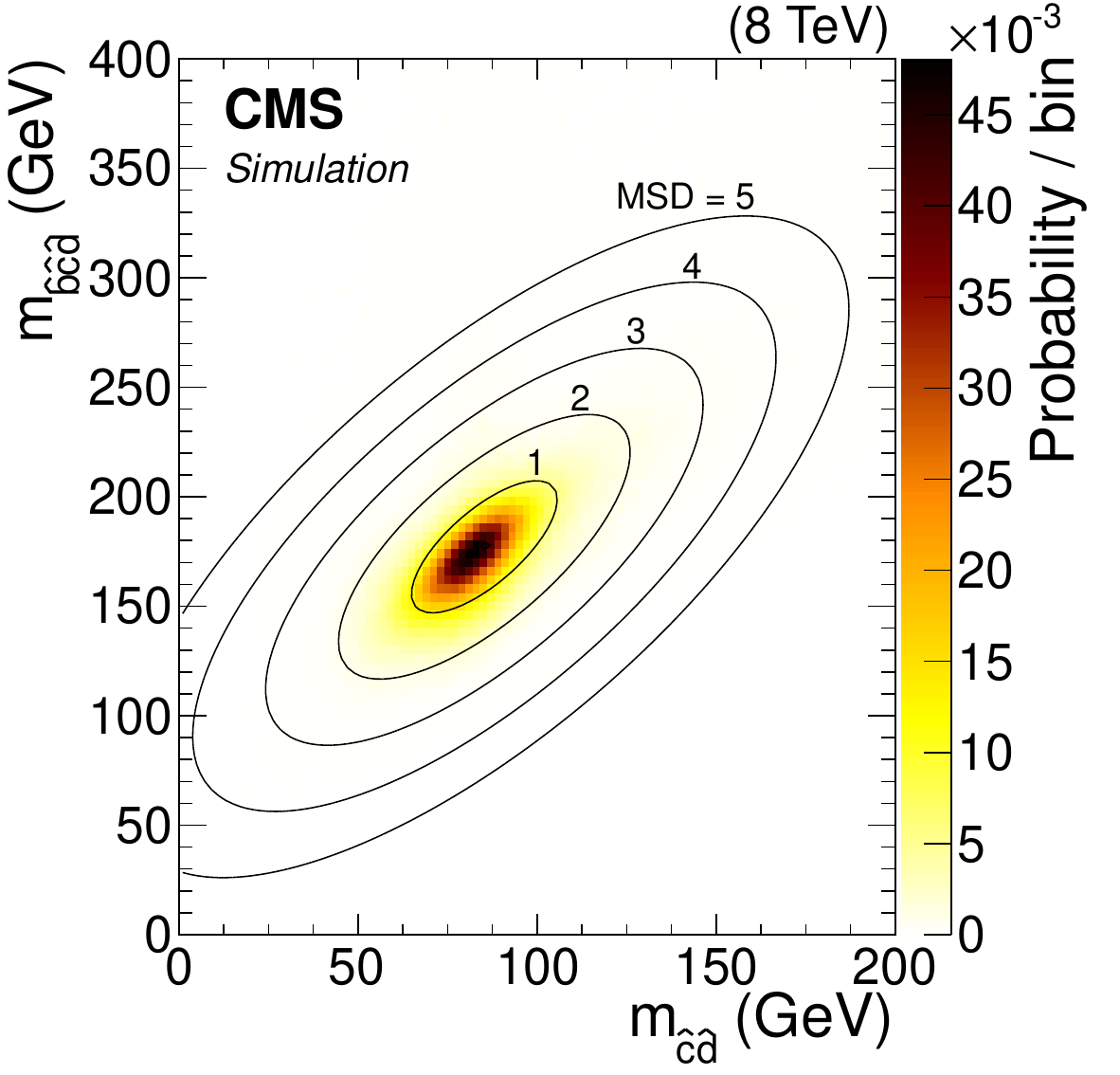}\\
  \includegraphics[width=0.4\linewidth]{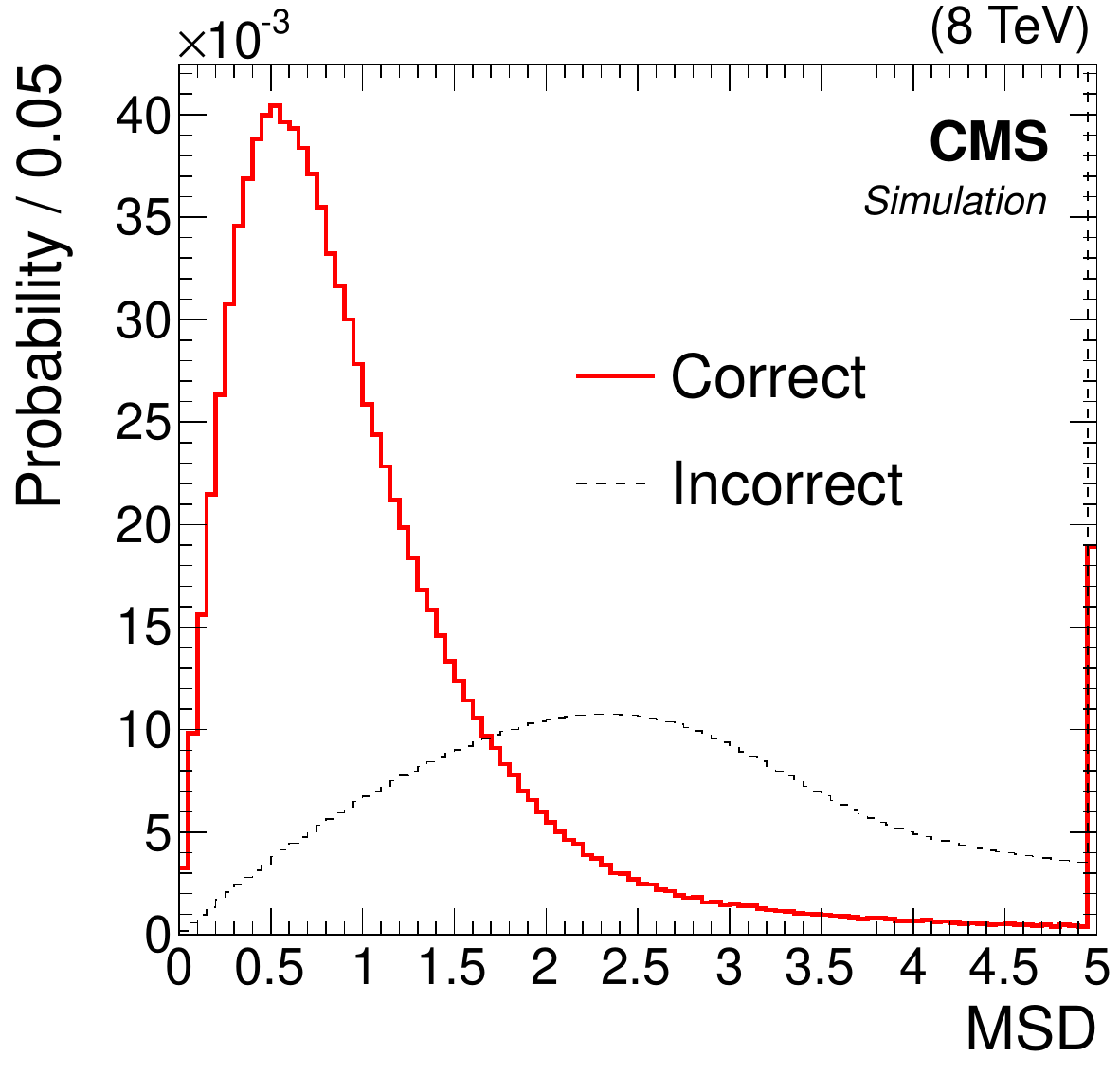}
  \qquad
  \includegraphics[width=0.4\linewidth]{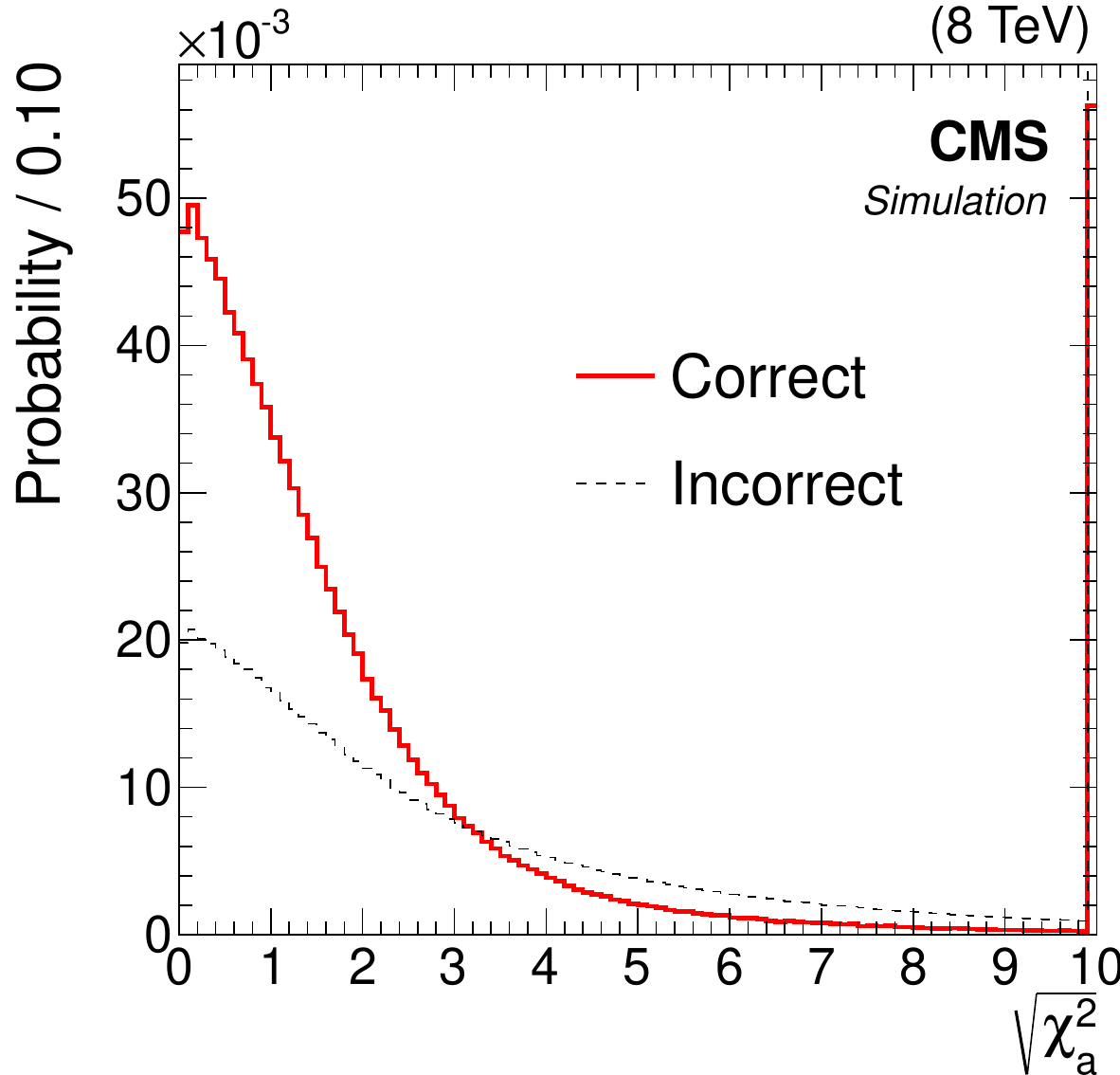}
  \includegraphics[width=0.4\linewidth]{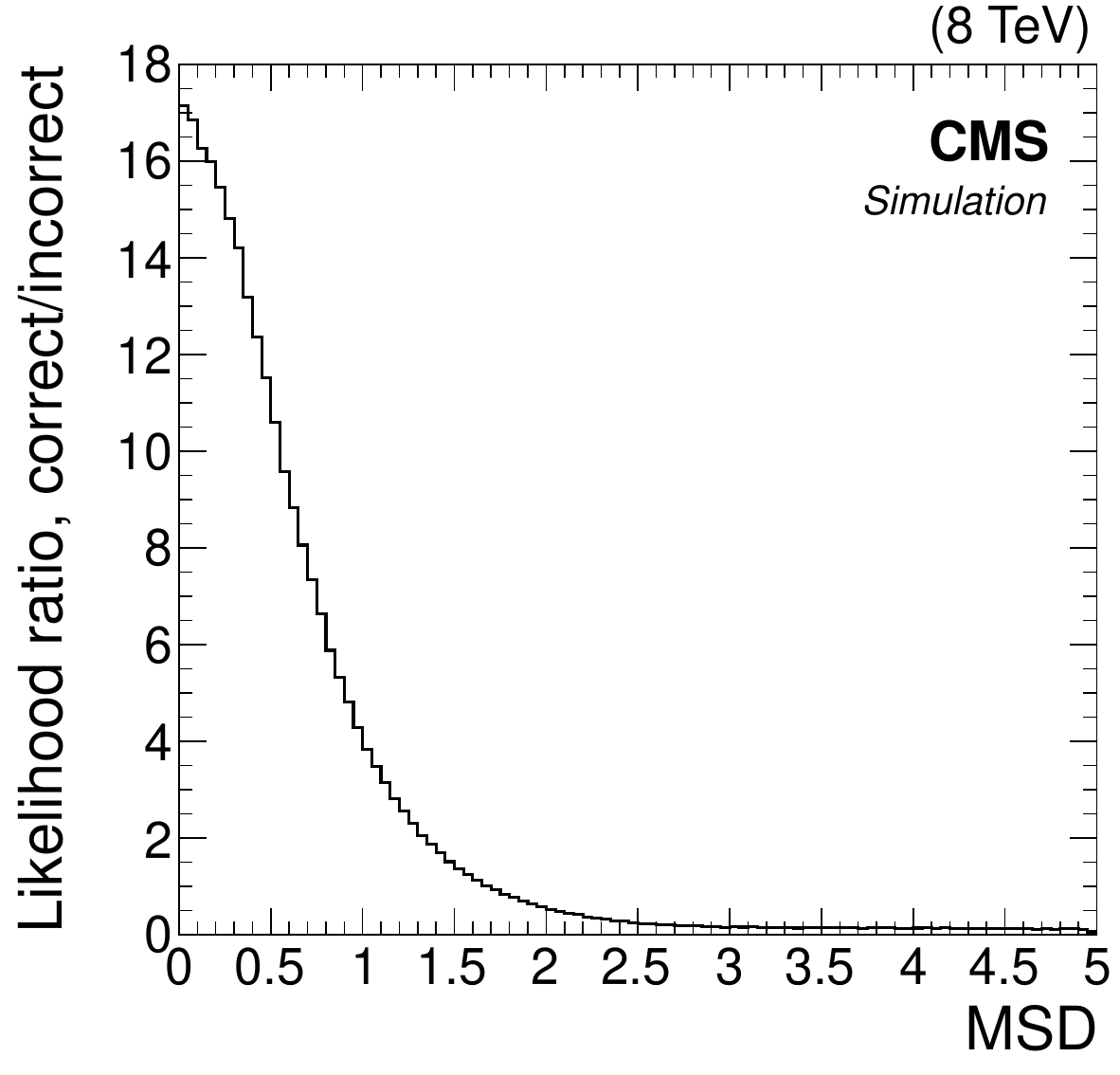}
  \qquad
  \includegraphics[width=0.4\linewidth]{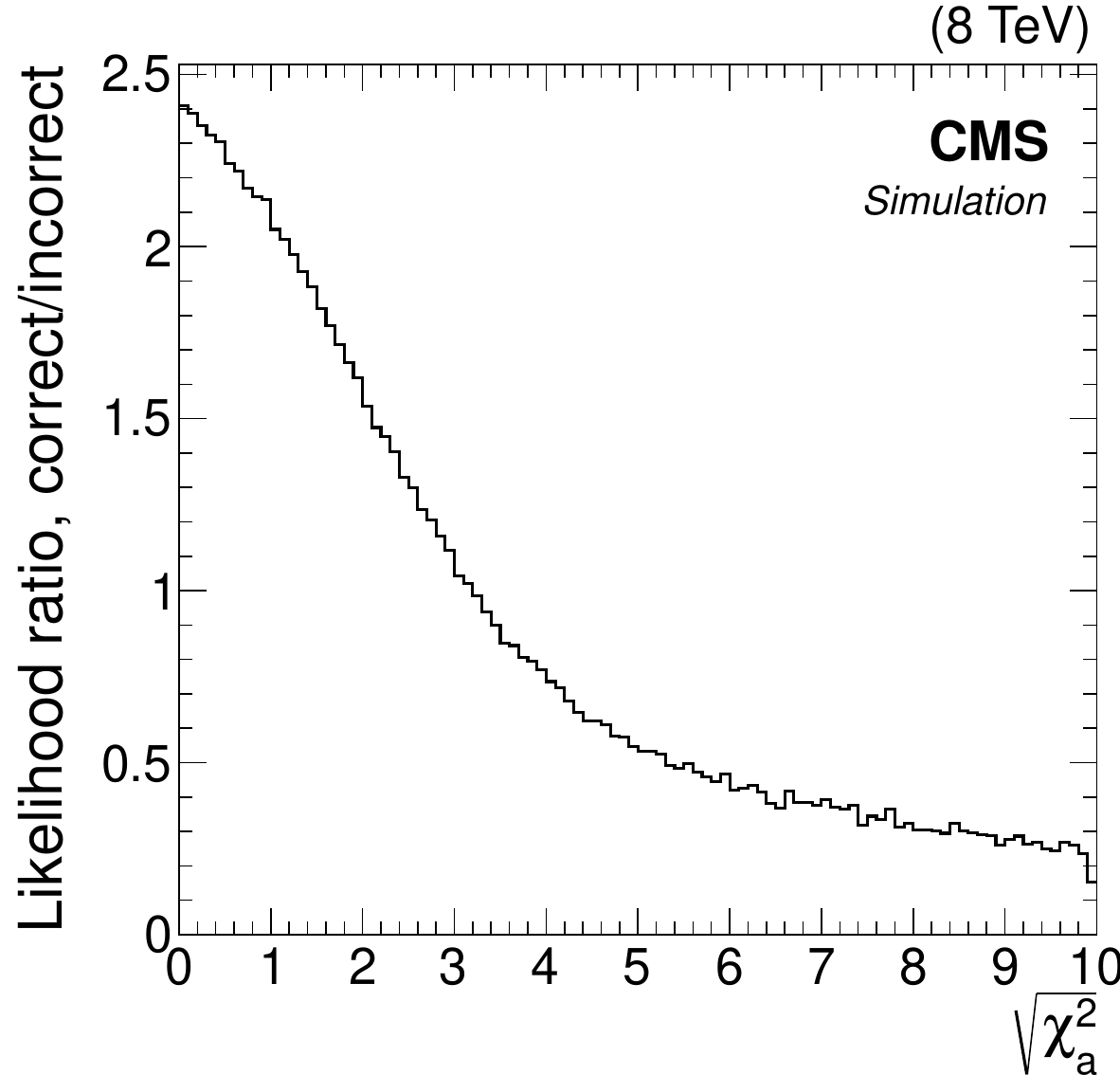}
  \caption{\label{likelihoods_ratios}
The two-dimensional probability density from simulation of jet invariant masses from
$\PW$ boson ($m_{\hat{c}\hat{d}}$) and top quark ($m_{\hat{b}\hat{c}\hat{d}}$) decay is shown (top),
along with contours in standard deviations (MSD) of the
corresponding Gaussian approximation.
Probability densities for correct and incorrect jet
assignments (middle) are shown (left) for MSD and (right) for
$\sqrt{\chi_a^2}$ of the leptonically decaying top quark
reconstruction.
The corresponding likelihood ratios are shown below.
}
\end{figure*}

The jet invariant masses associated with $\PQt\to\PQb\qqbar$ decays
are $m_{bcd}$ and $m_{cd}$, with parton-level jet corrections applied
based on the assignment.
Their two-dimensional probability distribution for correct assignments
is shown in \Fig{likelihoods_ratios}.
The mean and variance of this distribution are calculated after
removing the tail of the distribution, defined as the lowest-valued bins
which integrate to a 1\% probability, in order to find a Gaussian
approximation.
Contours of the approximation, in standard deviations, are also shown
in \Fig{likelihoods_ratios}.
The distance of a point from the center of this Gaussian function,
expressed in units of standard deviations, is denoted by ``mass standard
deviations'' (MSD).
Probability distributions in MSD for correct and incorrect
assignments, and their ratio $LR^{\mathrm{MSD}}$, are shown in
\Fig{likelihoods_ratios}.

The momentum of the neutrino associated with the leptonically decaying
top quark is calculated according to \Ref{Betchart:2013nba} using
\ptvecmiss and the four-momenta of the charged lepton and jet $a$.
Correct and incorrect assignments of jet $a$ are discriminated using the
test statistic
\begin{linenomath}
\begin{equation}
\label{chi2a}
\chi^2_a = \mathbf{d}^T\sigma^{-2}\mathbf{d},
\end{equation}
\end{linenomath}
where $\sigma^2$ is the covariance matrix for $\ptvecmiss$, derived from
the momentum uncertainties of the reconstructed objects in the event,
and $\mathbf{d}$ is the difference vector in the transverse plane
between $\ptvecmiss$ and the neutrino momentum solution.
The distributions of the square root of $\chi^2_a$ for correct and
incorrect assignments of jet $a$, and their ratio $LR^\chi$, are shown
in \Fig{likelihoods_ratios}.

Of the selected $\ttbar$ events, about half contain reconstructed jets
corresponding to all four \ttbar decay partons.
In about $60\%$ of those events, the assignment with the maximum likelihood
is also the correct assignment.

\subsubsection{Kinematic fitting procedure}

The energy resolution of jets corresponding to the most probable
assignment can be improved beyond the intrinsic resolution of the CMS
detector using the constraints from the masses of the top quark and $\PW$
boson.
These constraints are applied in two stages.
First, jet four-momenta $p_i$ are scaled to $\hat{p}_i = (1+\delta_i)p_i$
with the free parameters $\delta_i$, for $i$ equal to $b$, $c$, or
$d$, in the minimization of the test statistic
\begin{linenomath}
\begin{equation}
  \label{chi2had}
  \chi^2_{bcd} =
  \left(\frac{m_\PW-\hat{m}_{cd}}{\Gamma_\PW/2}\right)^2 +
  \left(\frac{m_\PQt-\hat{m}_{bcd}}{\Gamma_\PQt/2}\right)^2 +
  \sum_{i=bcd}\left(\frac{\delta_i}{r_i}\right)^2.
\end{equation}
\end{linenomath}
Here, $r_i$ are the $\pt$- and $\eta$-dependent relative jet energy
resolutions $\sigma_E/E$, and $\hat{m}_{cd}$ and $\hat{m}_{bcd}$ are
the invariant masses calculated with the scaled jet four-momenta.
The mass and width parameters used for the $\PW$ boson and top quark
are: $m_\PW=80.4\GeV$; $m_\PQt=172.0\GeV$; $\Gamma_\PW=2\GeV$;
and $\Gamma_\PQt=13\GeV$.
The values of $\Gamma_\PQt$ and $\Gamma_W$ represent the empirical
resolution of the reconstructed particle masses for a single event,
rather than the natural particle widths.
The momentum and energy of the top quark that decays into jets
are given by $\sum_{\{bcd\}}\hat{p}_i$.
In the second stage, the four-momentum of jet $a$ is scaled to
$\hat{p}_a = (1+\delta_a)p_a$ with the free parameter $\delta_a$, to
minimize the test statistic $\chi^2_a$ from \Eq{chi2a}.
At each step of this minimization, $\chi^2_a$ is calculated with the
charged-lepton four-momentum, the candidate $\hat{p}_a$, and $\ptvecmiss$
corrected for the scaling of the $a,\,b,\,c,$ and $d$ jets.
The uncertainty in the corrected $\ptvecmiss$ is reduced from that of the
nominal reconstruction by removing a portion of the uncertainty corresponding to
the energies of the $a,b,c$, and $d$ jets.
The neutrino momentum associated with the minimized $\chi^2_a$ is
summed with the corresponding $\hat{p}_a$ and the charged lepton
four-momentum to find the energy and momentum of the leptonically
decaying top quark.

\subsection{Discrimination among three populations}

\begin{figure}[htb]
  \begin{center}
  \includegraphics[width=0.35\textwidth]{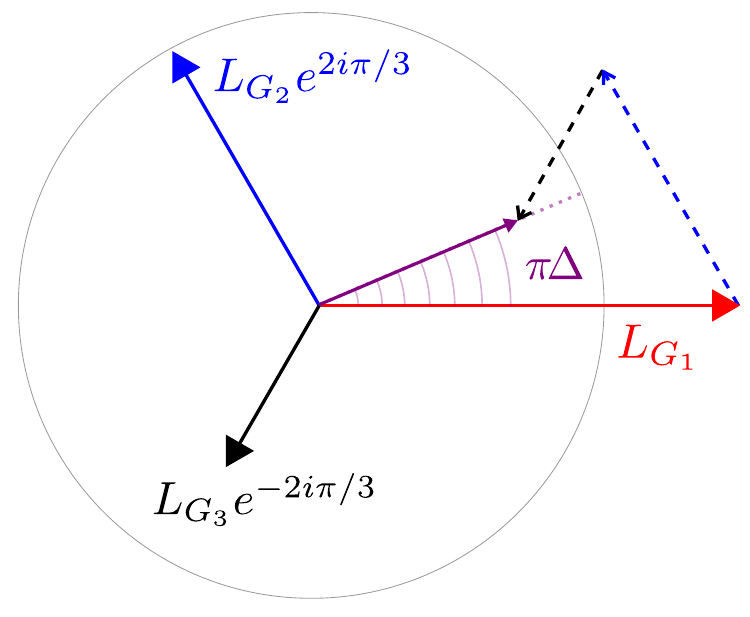}
  \end{center}
  \caption{\label{trid_construction} The angle $\pi\triD$ of the
    resultant sum of three vectors spaced at equal angles, in which
    the magnitude of each is the likelihood of the respective
    population.  The dashed arrows are translations of the $\re^{2i\pi/3}$ and
    $\re^{-2i\pi/3}$ vectors which illustrate the construction of the sum.
    The circle is shown to indicate the relative scale.
}
\end{figure}

To measure the sample composition in the data after the event
selection, we construct a likelihood discriminant designed to
distinguish among populations of events from three leading processes:
$\ttbar$, \MJ, and \Wj, denoted respectively by $G_1$, $G_2$, and $G_3$
in the following generalized construction.
As will be discussed in Section \ref{measurement_section}, the
contributions from \ST and \DY are constrained to those predicted by
their SM cross sections.
The likelihood that an event belongs to population $G$ is
$L_G=\prod_i\ell^G_i(V_i)$, where $\{V_i\}$ is a set of random
variables with probability densities $\ell_i^G$.
For independent $\{V_i\}$, the likelihood ratio $L_{G_2}/L_{G_1}$ is
more discriminating than any single constituent variable~\cite{welch1939}.
One can construct a likelihood-ratio-based discriminant
\begin{linenomath}
\begin{equation}
\label{tridiscriminant}
\providecommand{\Arg}{\mathrm{Arg}}
\triD = \Arg\left(L_{G_1} + \re^{2i\pi/3}L_{G_2} + \re^{-2i\pi/3}L_{G_3}\right) / \pi,
\end{equation}
\end{linenomath}
the principal value of which is bounded periodically on (-1,1] and is
symmetric under exchange of any two of the three populations.
Figure \ref{trid_construction} illustrates the
construction.
Populations $G_1,$ $G_2,$ and $G_3$ tend to concentrate at $\triD$ of
$0,$ $2/3,$ and $-2/3$, respectively.

Three observables are used to construct the likelihoods for the
discriminant.
The first is the transverse mass
$\MT = \sqrt{\smash[b]{2\lepT\ETmiss\,(1 -\cos\phi)}}$,
where $\lepT$ is the magnitude of the charged lepton \pt, $\phi$ is the azimuthal angle between the
charged lepton momentum and $\ptvecmiss$, and \ETmiss is the magnitude of $\ptvecmiss$.
The second is the probability from the MSD that at least one jet
assignment is the correct one, defined as $P_{\mathrm{MSD}} = \sum LR^{\mathrm{MSD}}_i /
\left(N_c+\sum LR^{\mathrm{MSD}}_i\right)$, where $N_c$ and $LR^{\mathrm{MSD}}_i$ are
defined in Section \ref{ttreco}.
The third is the probability from the CSV \PQb tagging discriminant
that at least one jet assignment is the correct one, defined as
\begin{linenomath}
\begin{equation}
  P_{\mathrm{CSV}} =
  \frac{\epsilon\sum L_i^{\mathrm{CSV}}}{
    \epsilon\sum L_i^{\mathrm{CSV}} +
    (1-\epsilon)N_c\prod_{j\in\{\text{jets}\}}\mathcal{N}(\beta_j)
  }\ ,
\end{equation}
\end{linenomath}
where $L_i^{\mathrm{CSV}}$ and $\mathcal{N}$ are defined in and before
\Eq{lCSV}, and the prior probability that at least one assignment is
correct is set to $\epsilon=0.05$.
A value of $\epsilon=0.05$ is chosen because it results in a more
balanced distribution of $P_{\mathrm{CSV}}$ than, for example, a flat prior with
$\epsilon=0.5$.
We found these observables to be highly discriminating and mostly
independent of each other.

\begin{figure*}[phtb]
  \centering
  \includegraphics[width=0.45\linewidth]{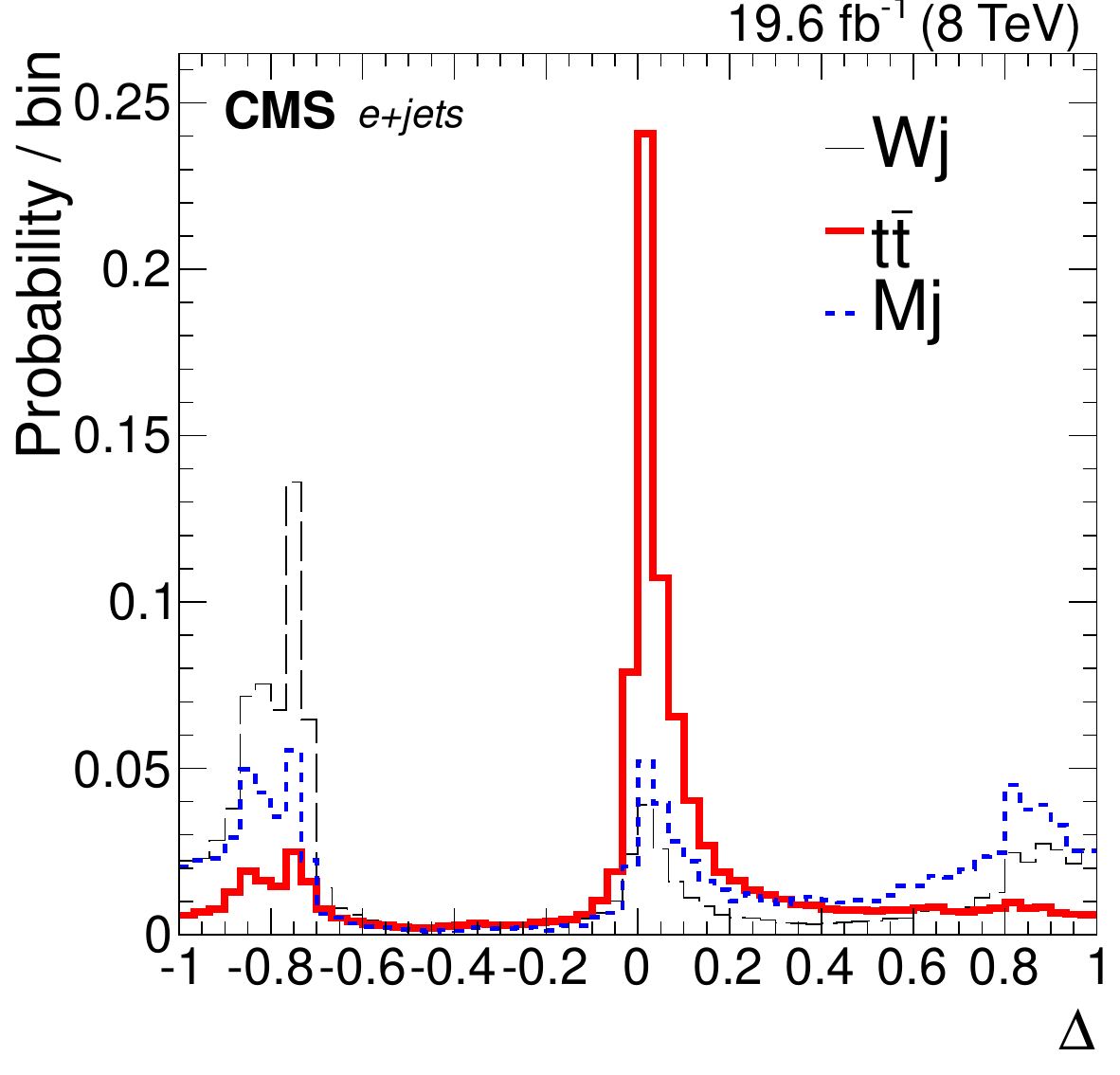}
  \qquad
  \includegraphics[width=0.45\linewidth]{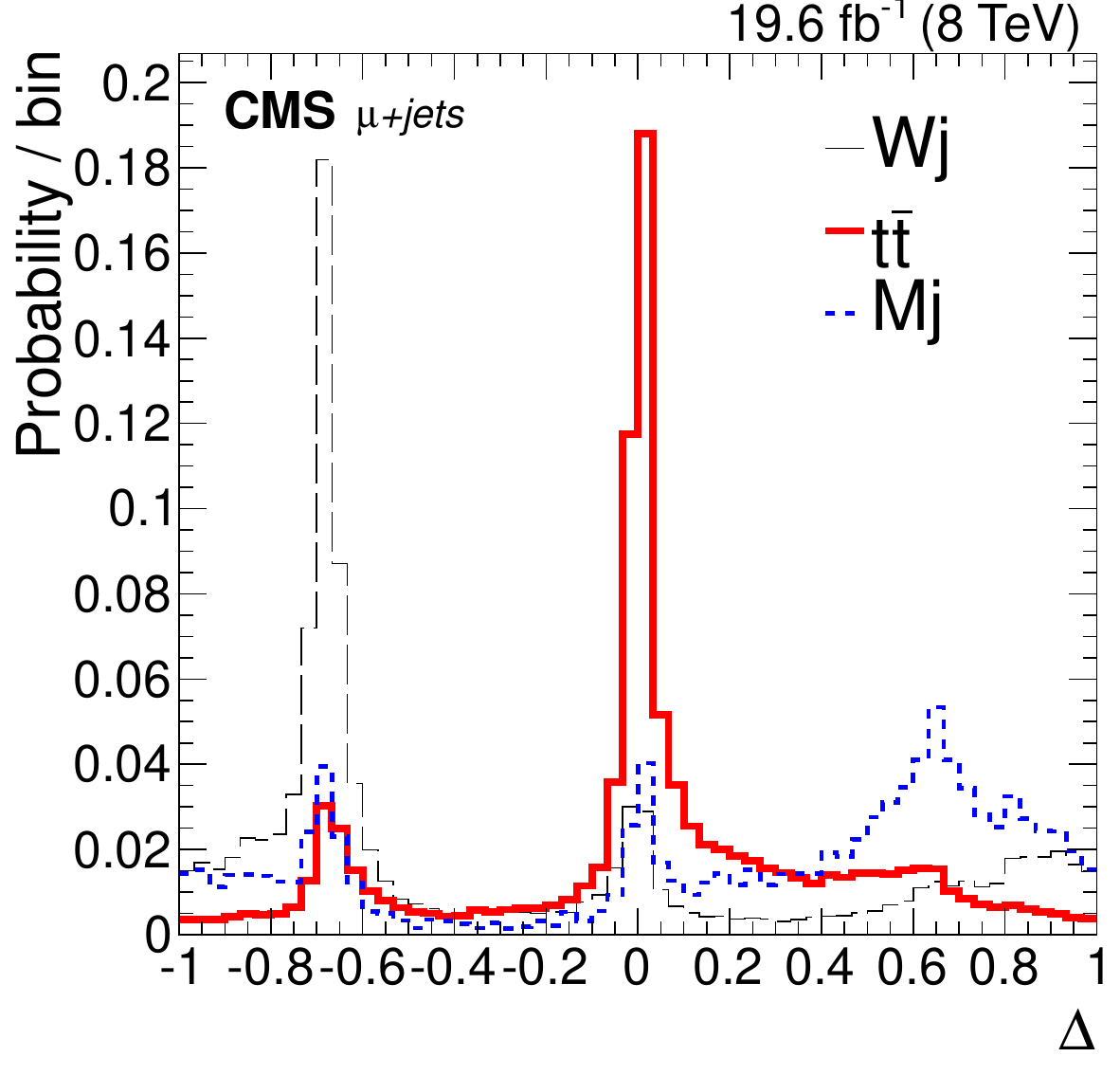}\\
   \includegraphics[width=0.3\linewidth]{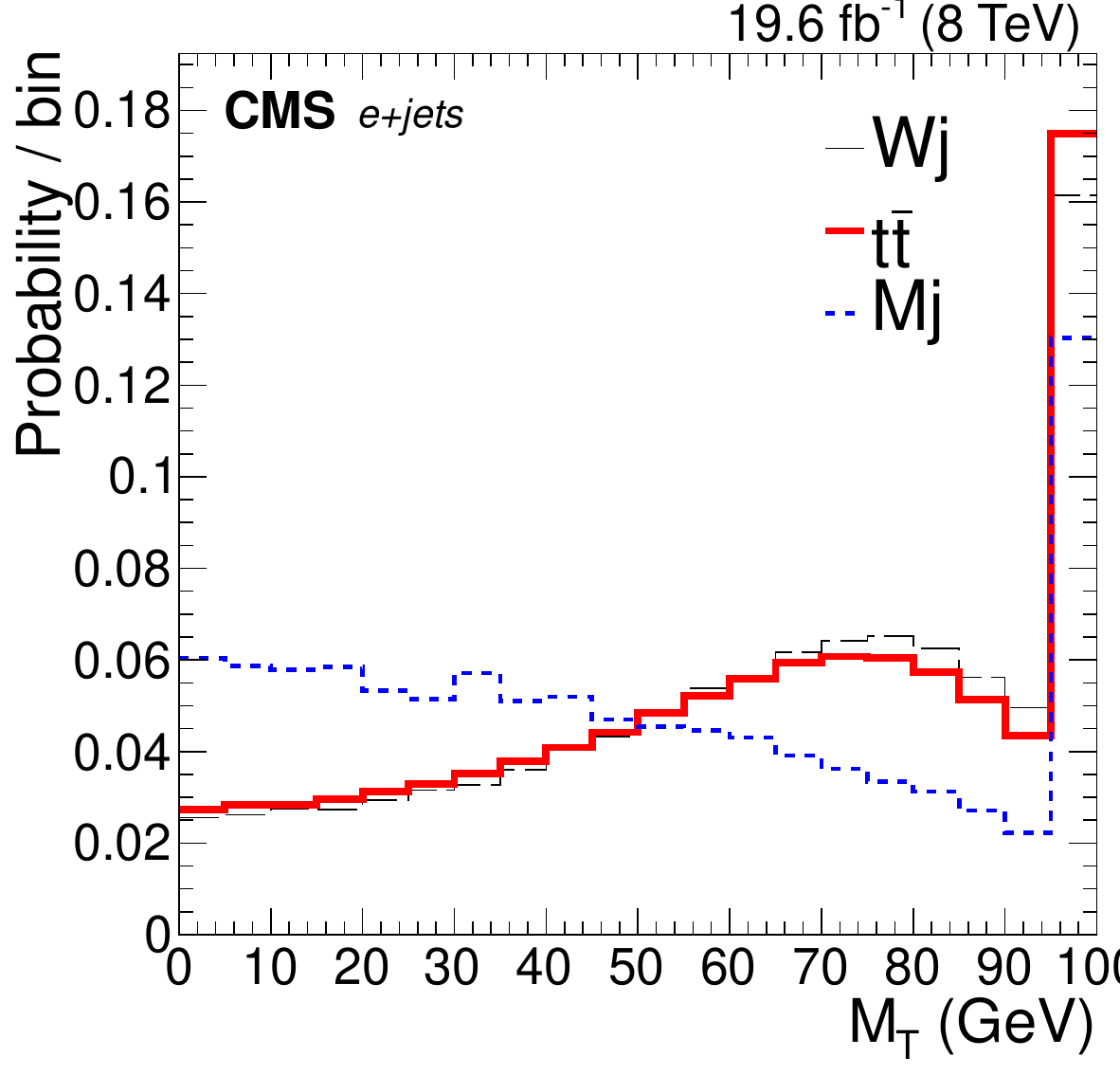}
  \quad
   \includegraphics[width=0.3\linewidth]{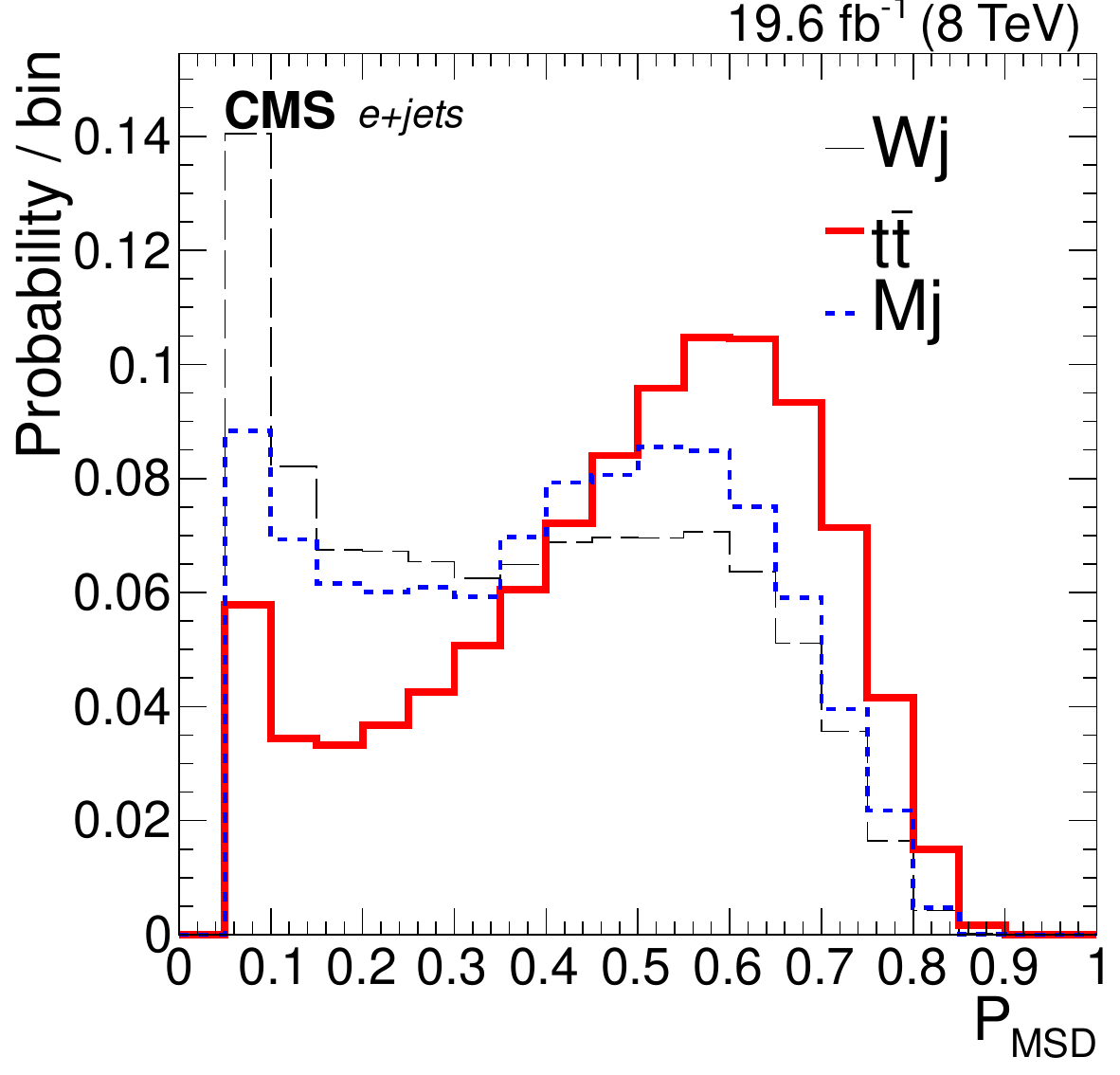}
  \quad
   \includegraphics[width=0.3\linewidth]{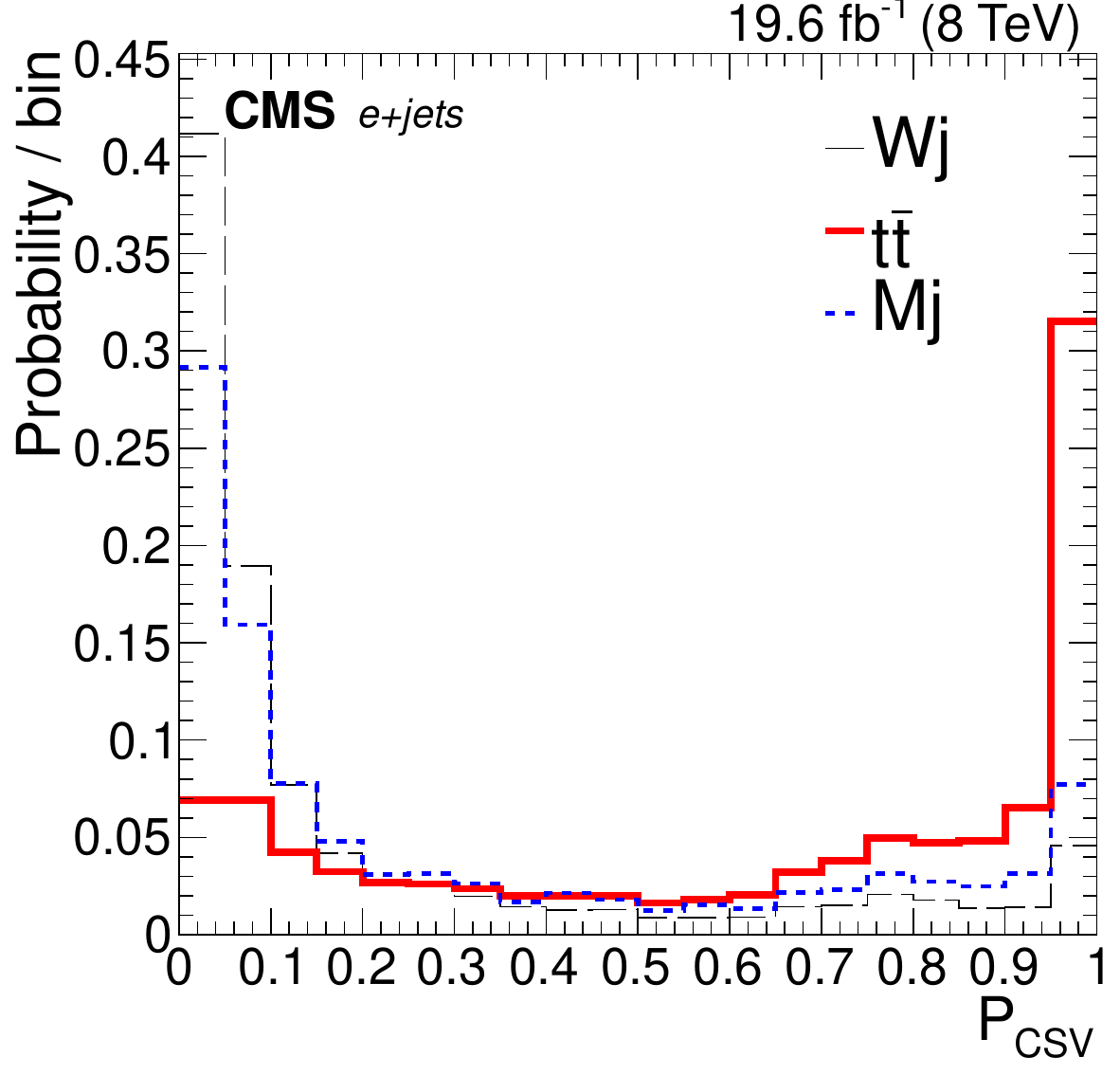}\\
   \includegraphics[width=0.3\linewidth]{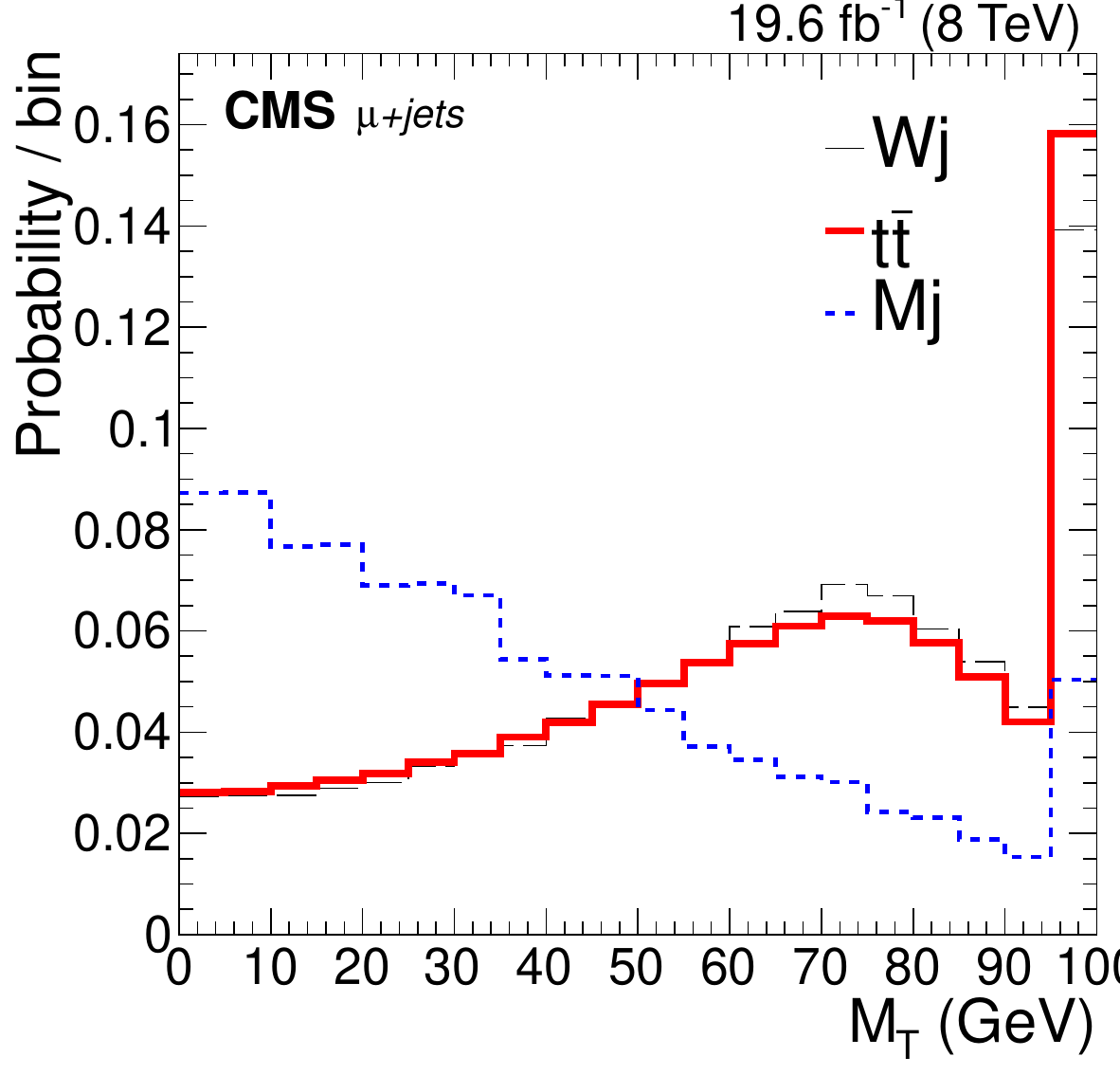}
  \quad
   \includegraphics[width=0.3\linewidth]{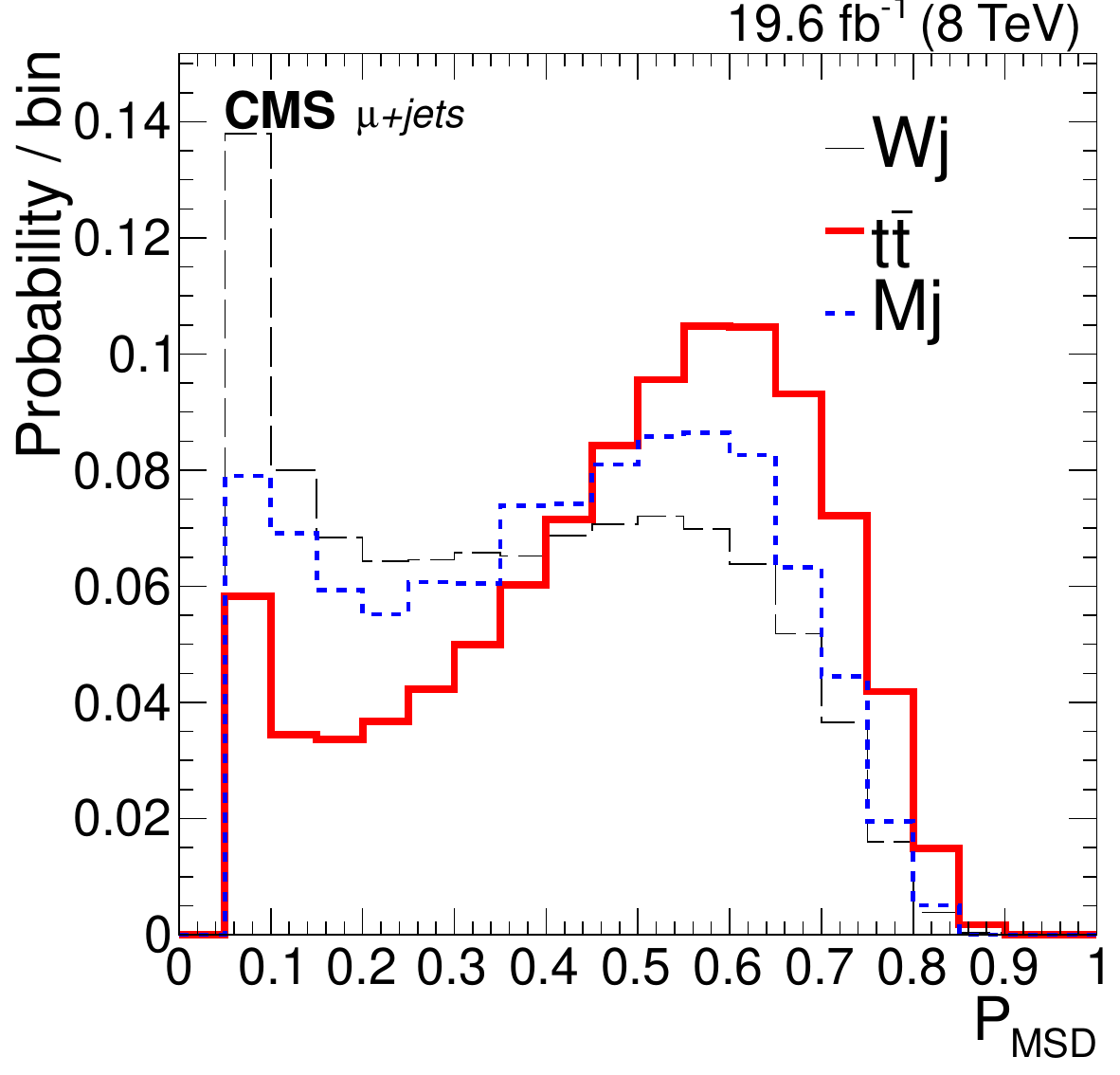}
  \quad
   \includegraphics[width=0.3\linewidth]{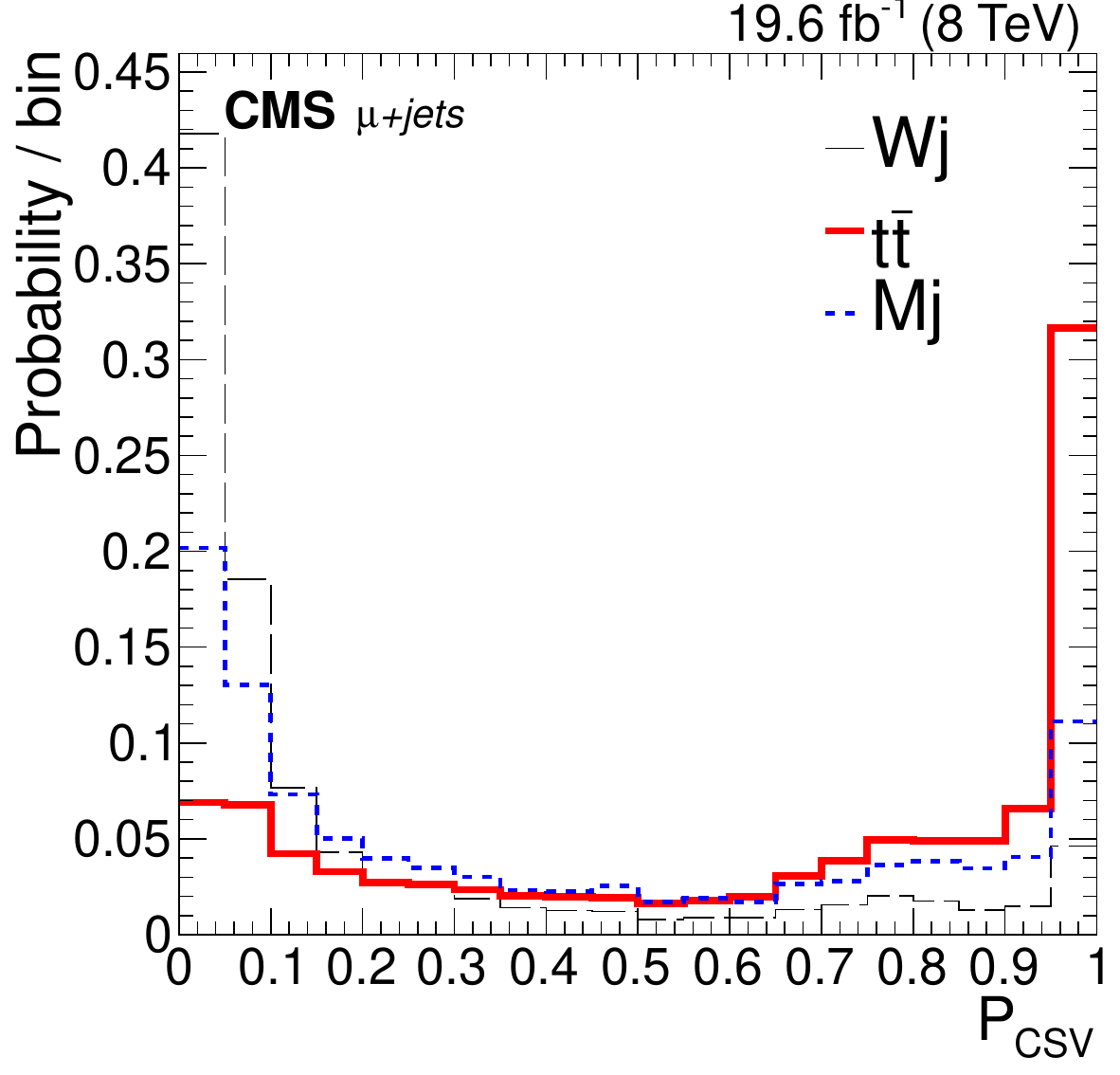}\\
  \caption{\label{triDpdfs}
The probability distribution of the discriminant $\triD$ for (top
left) selected \ejets\ events and (top right) selected
\mujets\ events, for the simulated \Wj and \ttbar populations, and for
the \MJ population, which is modeled from the sideband data with simulated
contributions subtracted.
The probability distributions in each observable used to construct the
discriminant are shown for
(middle) \ejets\ and (bottom) \mujets\ channels.
The overflow is included in the rightmost bin of the $\MT$ distributions.
}
\end{figure*}

The probability distribution for each population is shown as a function of
the discriminant and each of its input observables in
\Fig{triDpdfs}.
The \MJ probability distributions for the inputs are calculated using
fixed SM cross sections, as determined by the simulations, for the subtracted \ttbar and \Wj
contributions.

\section{Measurement procedure}
\label{measurement_section}
\label{templates_section}
\label{likelihood}

A two-stage maximum-likelihood fit is employed to sequentially measure
the sample composition, using the $\triD$ distribution, and the charge
asymmetry, using the $\XLreco$ distribution.

The sample composition is determined independently for each lepton
channel by fitting a model to the observed distribution $N_i^\ell$ in
the discriminant $\triD$.
Normalized five-bin templates in $\triD$ are constructed from the selected
events for each of the simulated processes, including \ttbar, \Wj,
\ST, and \DY, in both the signal and sideband regions.
The total number of events expected in each region from simulated
process $j$ is the product of the integrated luminosity $\lumi$, the
cross section $\sigma_j$, and the selection efficiency.
The selection efficiencies are taken directly from simulation.
Each cross section is parametrized by the relative change $\delta_j$
from the nominal value $\hat{\sigma}_j$.
The integrated luminosity is parametrized by the relative change
$\delta_\lumi$ from the measured central value.
The \MJ distribution in $\triD$ is determined at each iteration
of the fit by subtracting the sideband contributions of simulated
processes from the sideband region in data, and then rescaling this
distribution by a positive parameter $F^\ell_\MJ$.
The total number of expected events in each bin, $\lambda_i^\ell$, is
the sum of the expected contributions from the $\ttbar$,
\Wj, \MJ, \ST, and \DY processes.
Parameters $\delta_\lumi$, $\delta_\ST$, and $\delta_\DY$ are held
fixed to zero or to nonzero values when investigating systematic
uncertainties.
The sample composition is determined by finding values of the free
parameters $\{F^e_\MJ, F^\mu_\MJ, \delta_{\ttbar}, \delta_\Wj\}$ that
maximize the product of the Poisson likelihoods over the bins, given
observations $N^\ell_i$ and expectations $\lambda^\ell_i$.
The fit is implemented using \textsc{RooFit}~\cite{Verkerke:2003ir}.

\begin{figure}[htb]
  \centering
  \includegraphics[width=0.48\textwidth]{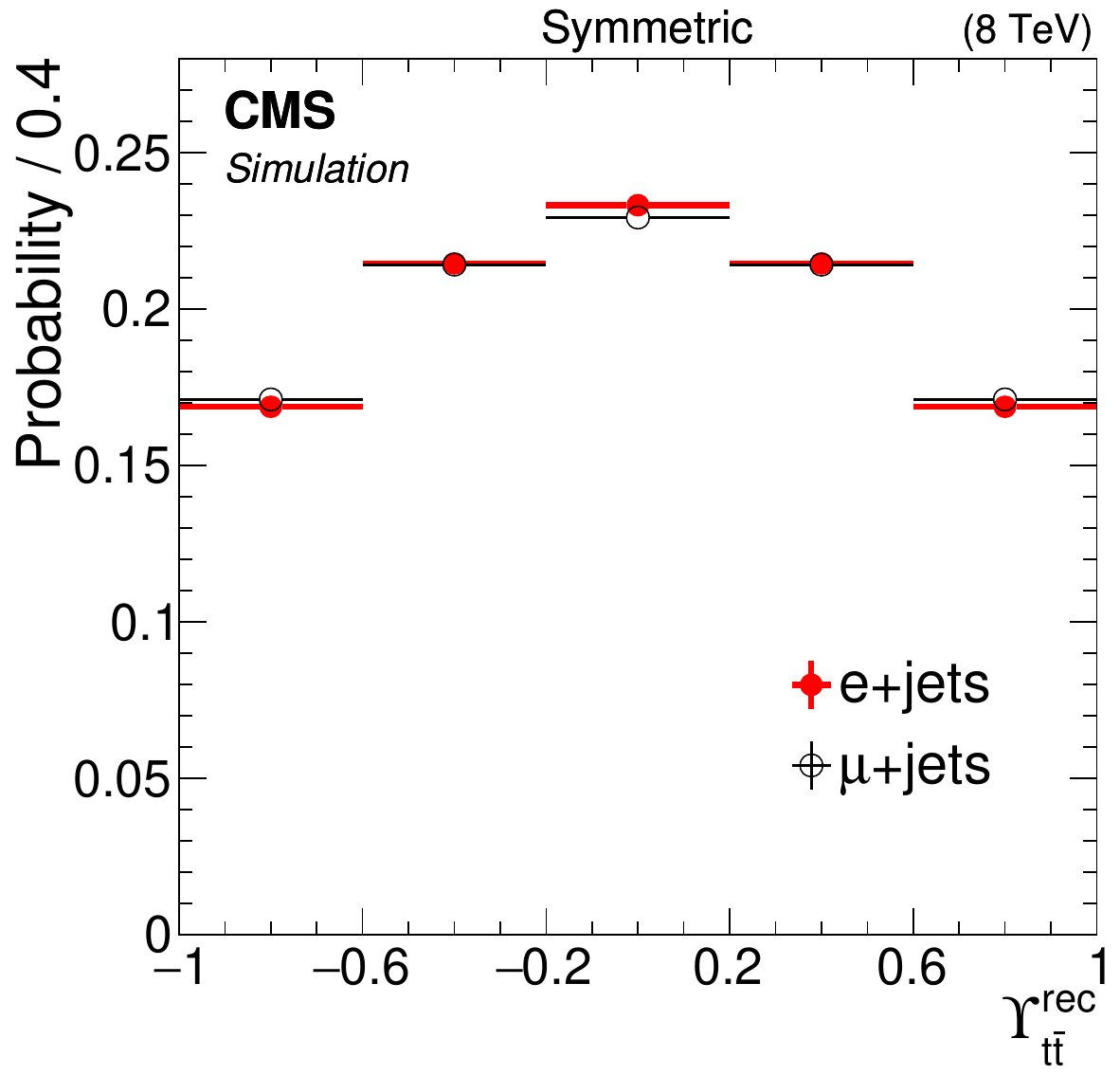}
  \includegraphics[width=0.48\textwidth]{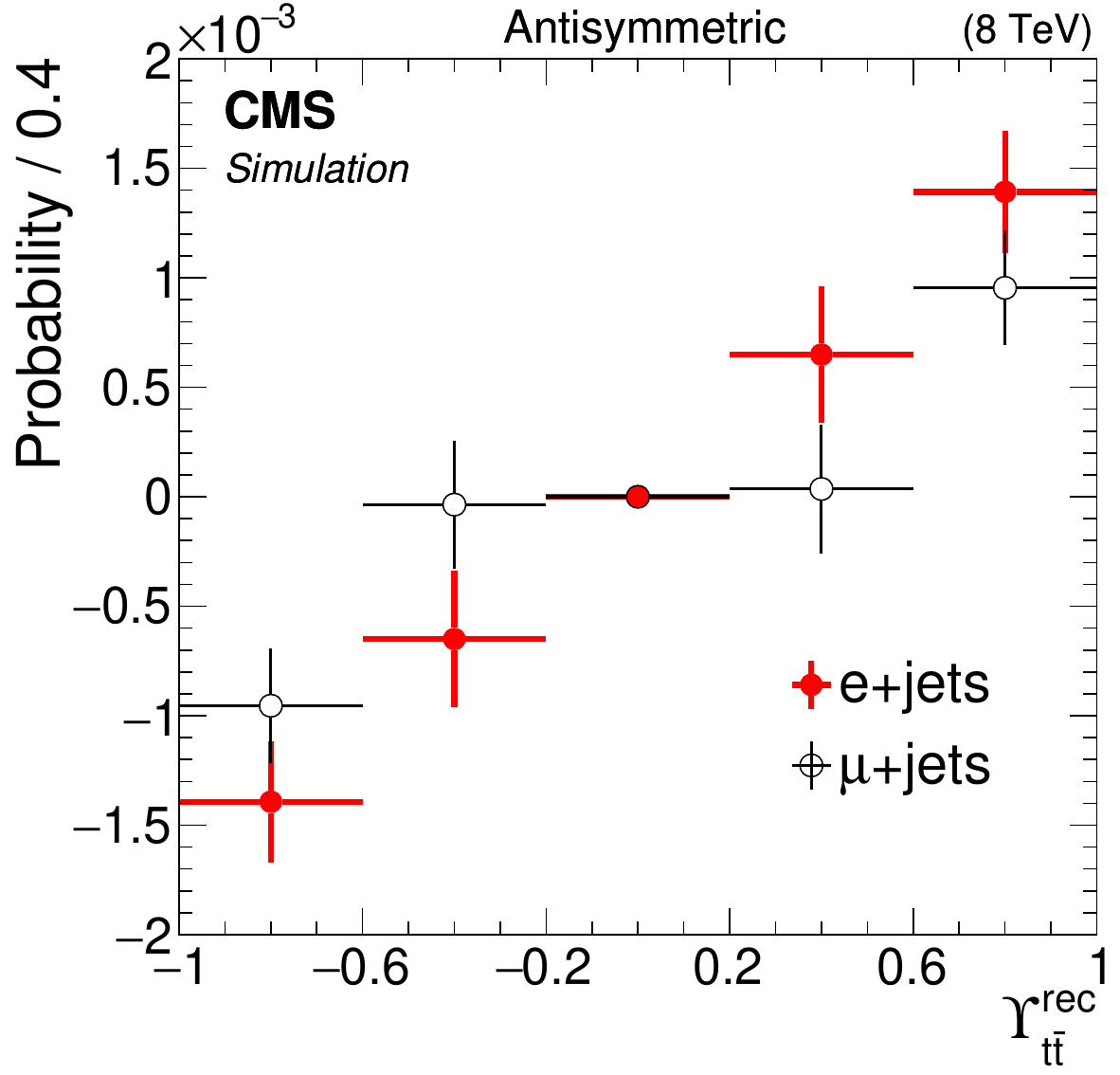}
  \caption{\label{templates} The (\cmsLeft) symmetric and (\cmsRight) antisymmetric components of the
    $\XLreco$ probability distribution for selected \ttbar simulation
    events in the \ejets\ and \mujets\ channels.
    The vertical bars show the statistical uncertainties, while the
    horizontal bars display the bin widths.
}
\end{figure}

The charge asymmetry is determined from a fit to the five-bin
distribution in $\XLreco$, based on the same model.
With the sample composition parameters held fixed, and following
\Eq{amplitude_reduced}, the \POWHEG \ttbar model is extended by
introducing a new free parameter $\alpha$ to provide changes in the
relative magnitudes of the symmetric and antisymmetric components of
$\XLreco$, shown in \Fig{templates}. The difference in shape of the
\ejets\ and \mujets\ templates is a result of the different rapidity
coverage between the two lepton flavors.
The modeled charge asymmetry is that of the \ttbar base model,
$\hat{A}_c^y$, scaled by $\alpha$,
\begin{linenomath}
\begin{equation}
  A_c^{y} = \alpha\hat{A}_c^{y}.
\end{equation}
\end{linenomath}
The charge asymmetry in the data is estimated by finding the value of
$\alpha$ that maximizes the product of the Poisson likelihoods over the bins.
The results from the independent measurements in both lepton channels are
combined before evaluating the systematic uncertainties.

\subsection{Performance and calibration}

\begin{figure}[htb]
  \centering
  \includegraphics[width=0.48\textwidth]{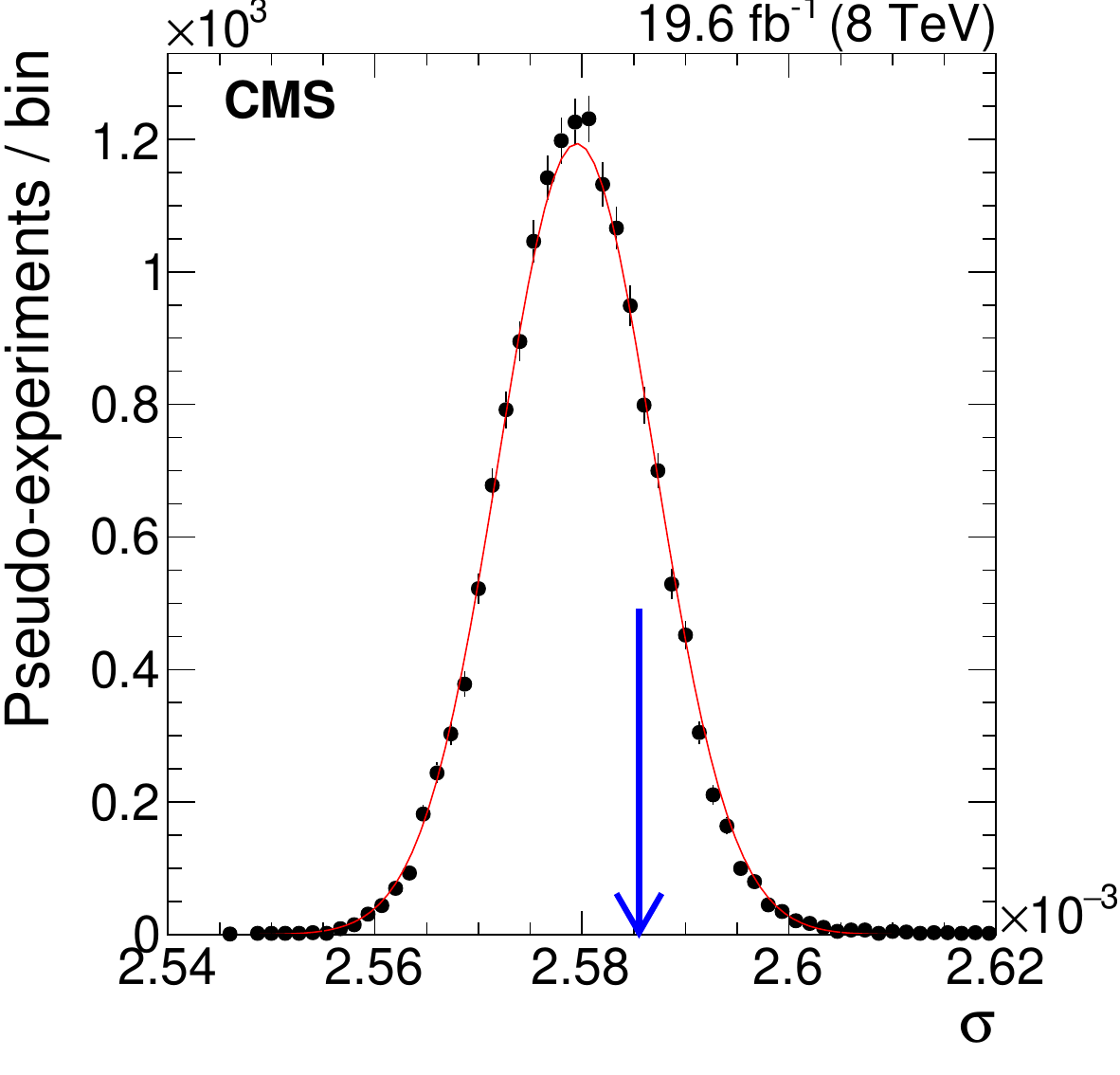}
  \caption{\label{uncertainty} The distribution of the statistical
    uncertainty in $A_c^y$ from measurements using pseudo-experiments, with
    an expected value of 0.258\%.  The statistical uncertainty
    extracted from the data is marked by the arrow.}
\end{figure}

The performance of the method is checked on simulated samples constructed using
\ttbar events based on the extended \POWHEG model as well as the
alternative \ttbar simulations described in Section
\ref{signal_templates}.
The extended \POWHEG model is checked using various values of the parameter
$\alpha$ by measuring pseudo-experiments generated with Poisson
variations of the best-fit model, mimicking fluctuations expected
in data.
The statistical uncertainty measured in 68\% of the pseudo-experiments is
greater than the absolute difference between the measured and expected
values.
The distribution in statistical uncertainty in $A_c^y$, with an expected value
of 0.258\%, is shown in \Fig{uncertainty}.

\begin{figure}[htb]
  \centering
  \includegraphics[width=\cmsFigWidth]{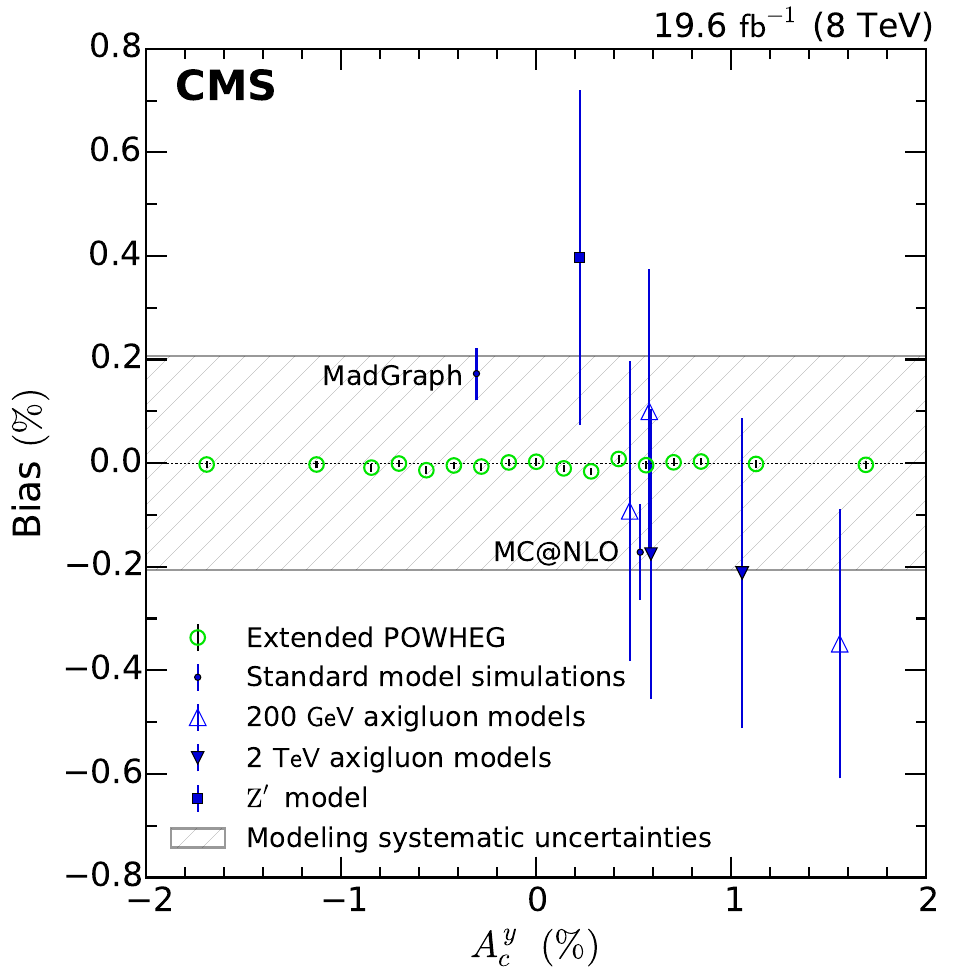}
  \caption{\label{bias_plot}
The bias in the measured charge asymmetry for SM simulations and  alternative $\ttbar$
models, based on extended \POWHEG SM templates, versus the charge
asymmetry in each sample.
The beyond-SM samples are \MADGRAPH simulations of \PZpr bosons and
axigluons with masses of 200\GeV and 2\TeV.
Uncertainty in the bias of the extended \POWHEG model is dominated by the
number of pseudo-experiments used, while the uncertainty in the bias
of each alternative model is dominated by the statistical uncertainty in the sample.
The hatched area shows the systematic uncertainty in the measurement
of $A_c^y$ from sources related to the modeling, including simulation
statistical uncertainty, renormalization and factorization scales, choice of \ttbar
generator, top quark mass, and PDFs.
}
\end{figure}

The alternative \ttbar simulations are checked using
pseudo-experiments with the sample composition of the measured data,
constructed with fixed background and Poisson-varied signal templates,
to find the uncertainty from the sample statistics of each
alternative model.
Identical background samples are used in constructing the pseudo-data
and in constructing the measurement model, so statistical uncertainty
in the background samples does not contribute to uncertainty in the
calibration.
Figure \ref{bias_plot} shows the difference between the expected
measurement and the input charge asymmetries, or the bias, for each
model.
The bias for the extended \POWHEG models is negligible.
The bias of the method when applied to samples produced using the
SM-based generators \MADGRAPH and \MCATNLO is compatible with the
systematic uncertainty in $A_c^y$ assigned to model-related sources, represented
by the shaded band in the plot.
Model-related systematic uncertainty sources consist of simulation
statistics, modeling of \ttbar production, PDFs, and renormalization
and factorization scales.
Similar calibrations of the beyond-SM alternatives of \ttbar production
considered in this study all show biases statistically compatible with
zero.

\subsection{Systematic uncertainties}

Systematic uncertainties in $\alpha$ are investigated after the
statistical combination of the two channels by repeating the
measurement with variations in the parameters or the distributions.
The second stage of the fit is repeated with sample composition
parameters varied independently to the upper and lower bounds of their
68\% confidence intervals.
Parameters for the integrated luminosity and the \ST and \DY cross
sections are varied similarly, but both fit stages are repeated.
The effects of statistical uncertainty in the sideband distributions of the data
and the simulations are investigated with ensembles of
alternative templates, generated by varying the originals according to
Poisson statistics.
Uncertainty in the jet energy scale and jet energy resolutions are
investigated by repeating the reconstruction using rescaled jet
energies, according to the \pt and $\eta$ of each jet.
Likewise, the modeling of the \PQb tagging discriminator is varied
by repeating the reconstruction with scaled discriminant values.
The PDFs are varied by event reweighting of the \ttbar templates to the
90\% confidence limits of each of the 26 CT10 eigenvectors and the strong
coupling parameter, independently; we chose to use this method rather 
than the widely used PDF4LHC prescription~\cite{pdf4lhc} since the former is
sensitive to the possibility of a strong correlation between the
antisymmetric component of the $\XLreco$ distribution and any
eigenvector, while varying the distribution to the minimum and maximum
of the uncertainty envelope is not.
Uncertainty from the modeling of \ttbar production is estimated by
measuring the data using extended \MCATNLO templates rather than
the extended \POWHEG templates, and varying the top quark mass by ${\pm}0.9\GeV$.
The factorization and renormalization scales are varied by
substituting distinct samples for the \ttbar templates, described in
Section \ref{signal_templates}.
The heavy-flavor content of \Wj events is varied by adding or
subtracting 20\%~\cite{Campbell:2010ff} of the expected contribution of a distinct
$\PW$+$\bbbar$ sample to the expected \Wj templates.
Variations in distributions for the pileup multiplicity and the top
quark \pt, and variations in the trigger and identification
efficiencies for the charged leptons, are accomplished by event
reweighting.
The uncertainty in the shape of the \MJ templates is dominated by
the statistical uncertainty in the data sidebands; the \MJ
antisymmetric components are statistically compatible with zero
asymmetry, and no additional shape systematic is included beyond that
of the statistical shape uncertainty.

The magnitudes of the systematic uncertainties are given in
\Tab{list_systematics}.
The total systematic uncertainty of 0.33\% is comparable to the
statistical uncertainty in the measurement, and is dominated
by the statistical uncertainty in the shapes of the data sidebands.
\begin{table}[htb]
\centering
\topcaption{\label{list_systematics} Uncertainty in the combined
measurement of $A_c^y$ from systematic sources,
ordered by decreasing magnitude.  }
\begin{scotch}{.l}
\multicolumn{1}{c}{(\%)}                                & Source of systematic uncertainty in $A_c^y$\\
\hline
               0.18 & Data sideband statistical uncertainty \\
               0.15  & Simulation statistical uncertainty \\
               0.14  & Jet energy scale \\
               0.14  & Renormalization and factorization scales \\
               0.073  & Modeling of \PQb tagging\\
               0.037  & $\sigma_\ST$ ($\sigma_{\PQt}+\sigma_{\PAQt}$) \\
               0.035  & Jet energy resolution \\
               0.026  & Modeling of pileup \\
               0.023  & $\PW\bbbar$ content \\
               0.021  & Ratio of \ST cross sections, $\sigma_{\PQt} / \sigma_{\PAQt}$ \\
               0.021  & Modeling of $\ttbar$ production\\
               0.018  & PDFs \\
               {<}0.010  & $\lumi$, $\sigma_\DY$, $\delta_\Wj$, trigger $\epsilon_\PGm$, $F^\Pe_{\MJ}$, $\delta_{\ttbar}$, $\alpha_s$ \\
               {<}0.001  & Trigger $\epsilon_\Pe$, $\pt^{\PQt}$, ID$_\Pe$, ID$_\PGm$, $F^\PGm_{\MJ}$ \\
               \hline
              0.33  & {Total} \\
\end{scotch}
\end{table}

\section{Results}
\label{results_section}

\begin{figure*}[htbp]
  \centering
  \includegraphics[width=\projwidth]{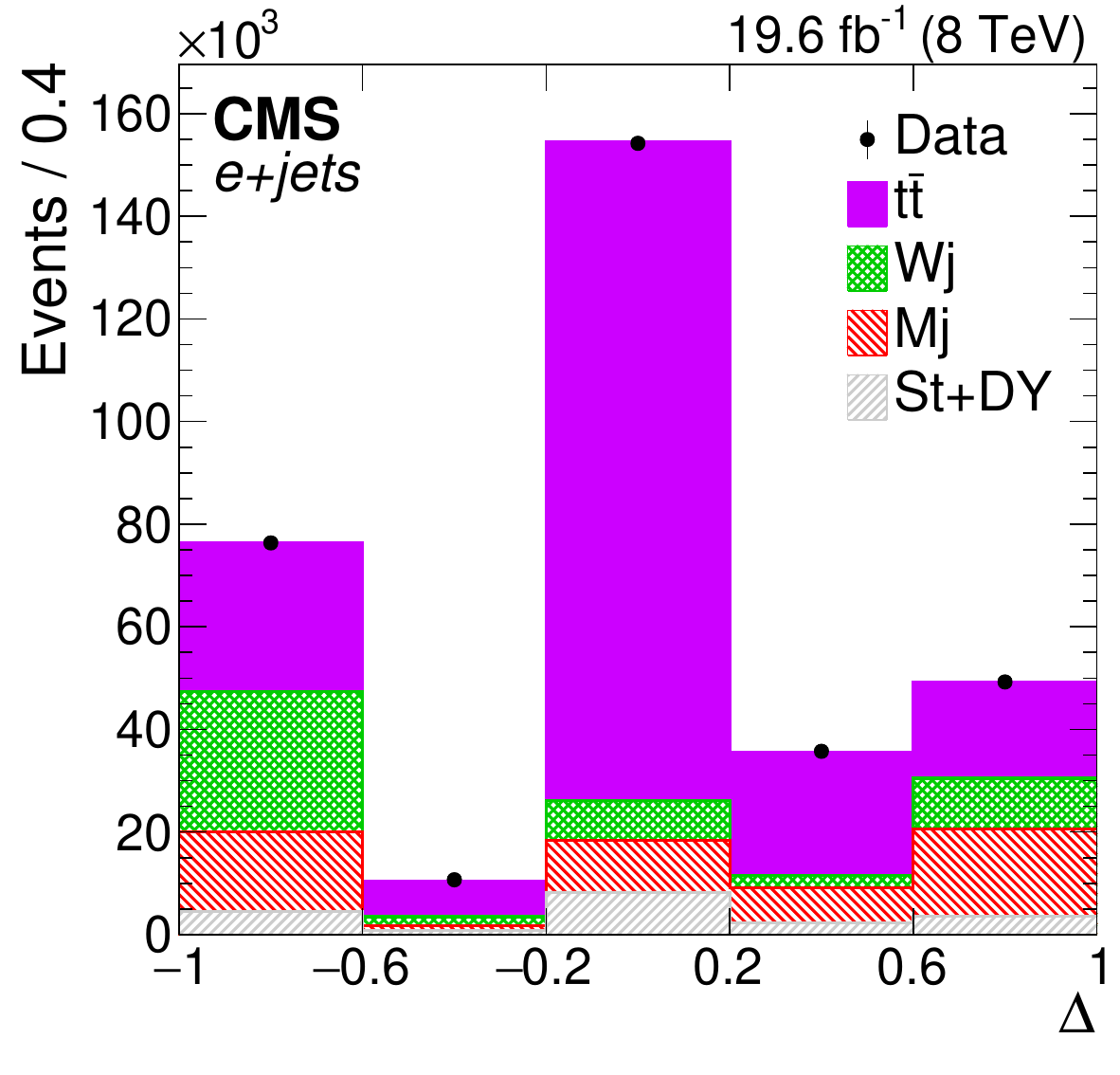}
  \quad
  \includegraphics[width=\projwidth]{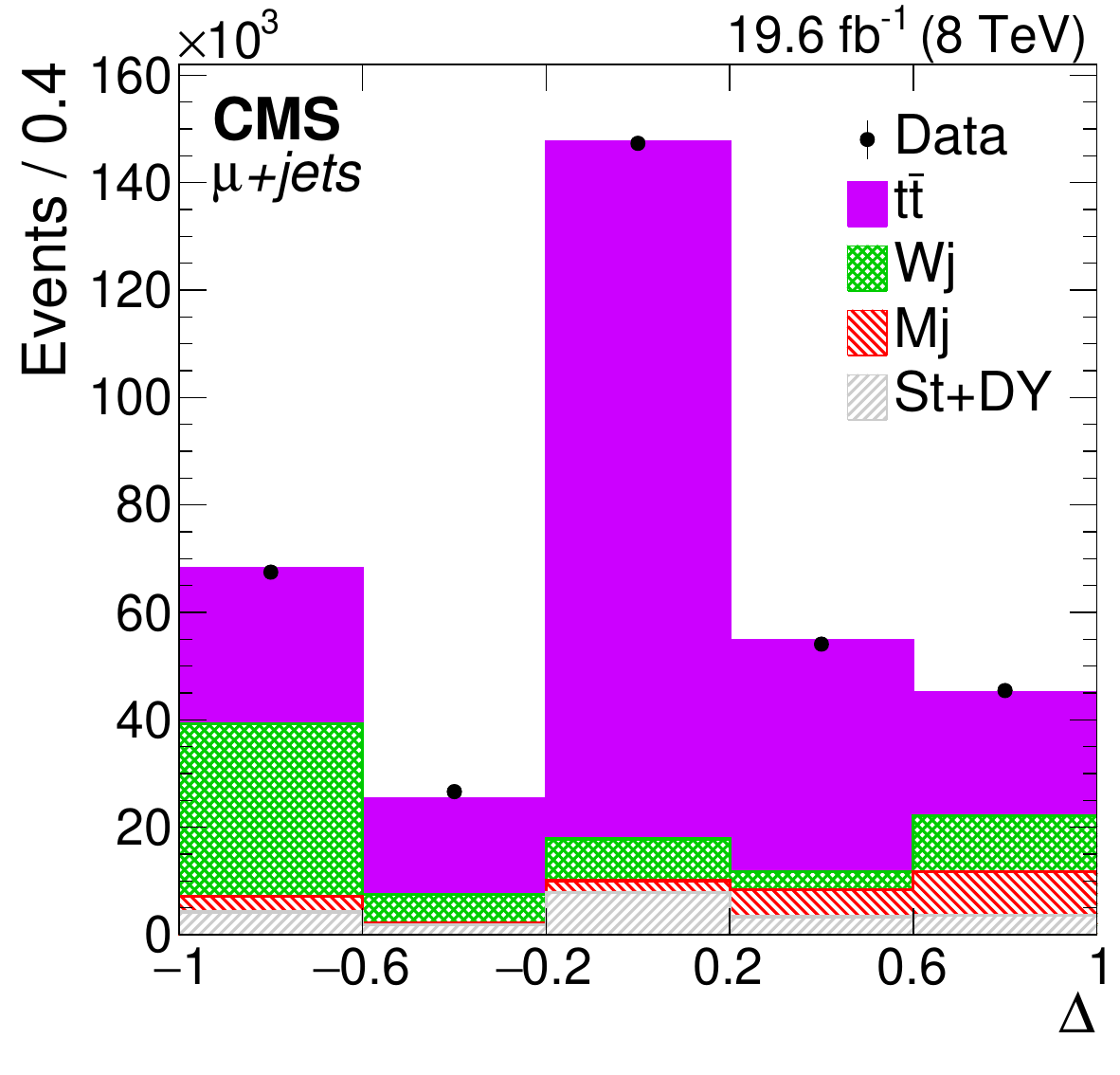}\\
  \includegraphics[width=\projwidth]{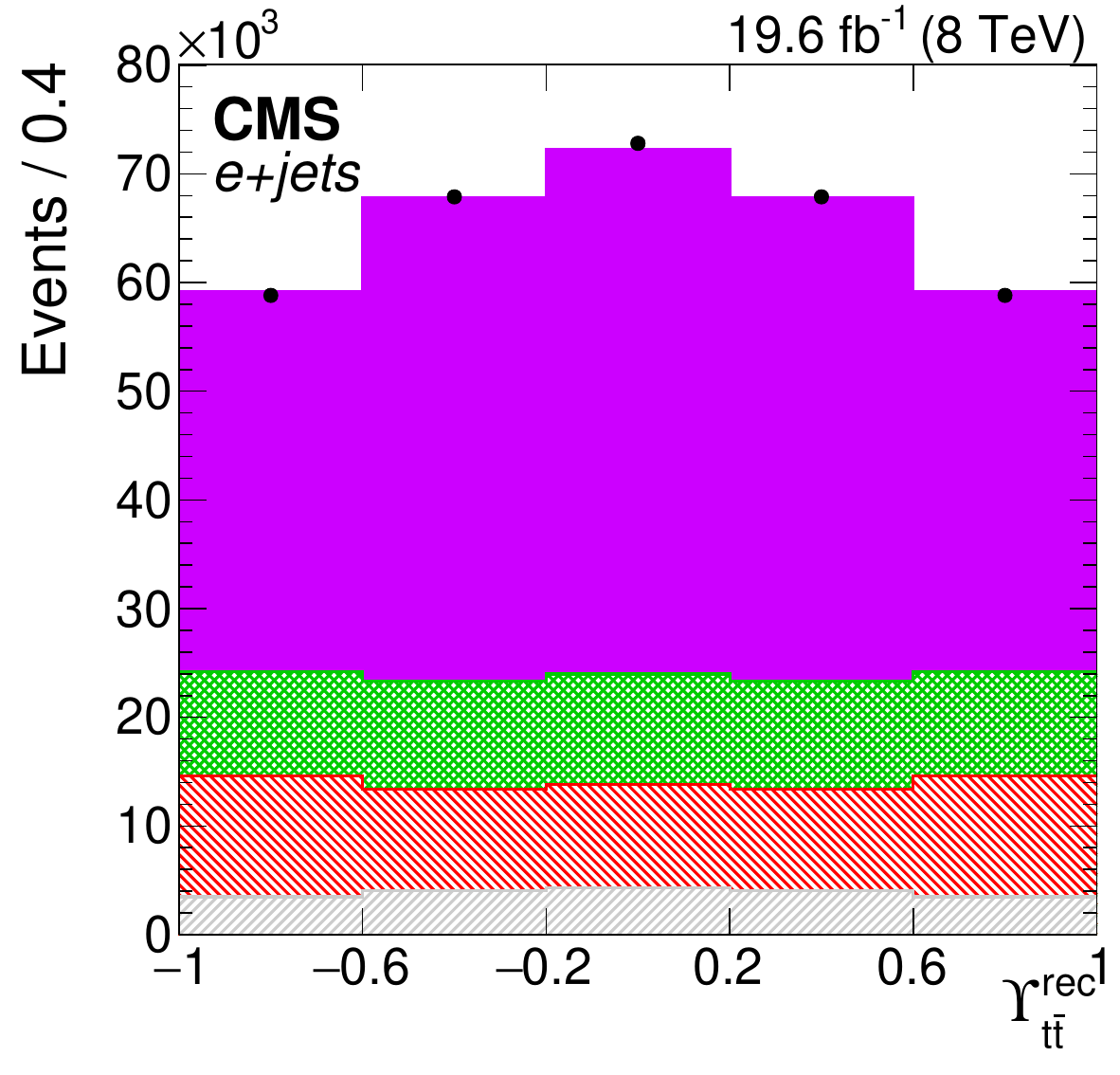}
  \quad
  \includegraphics[width=\projwidth]{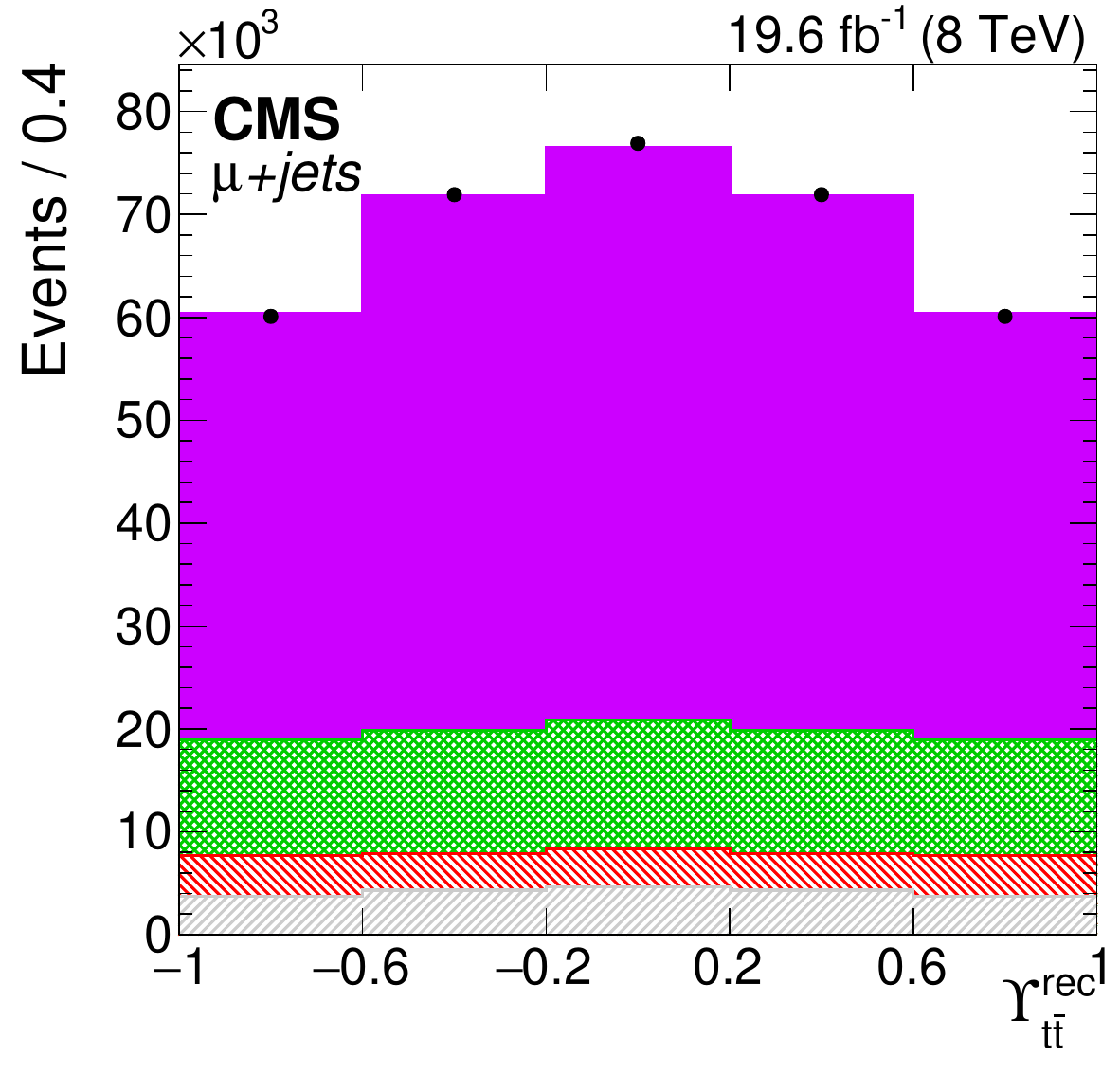}\\
  \includegraphics[width=\projwidth]{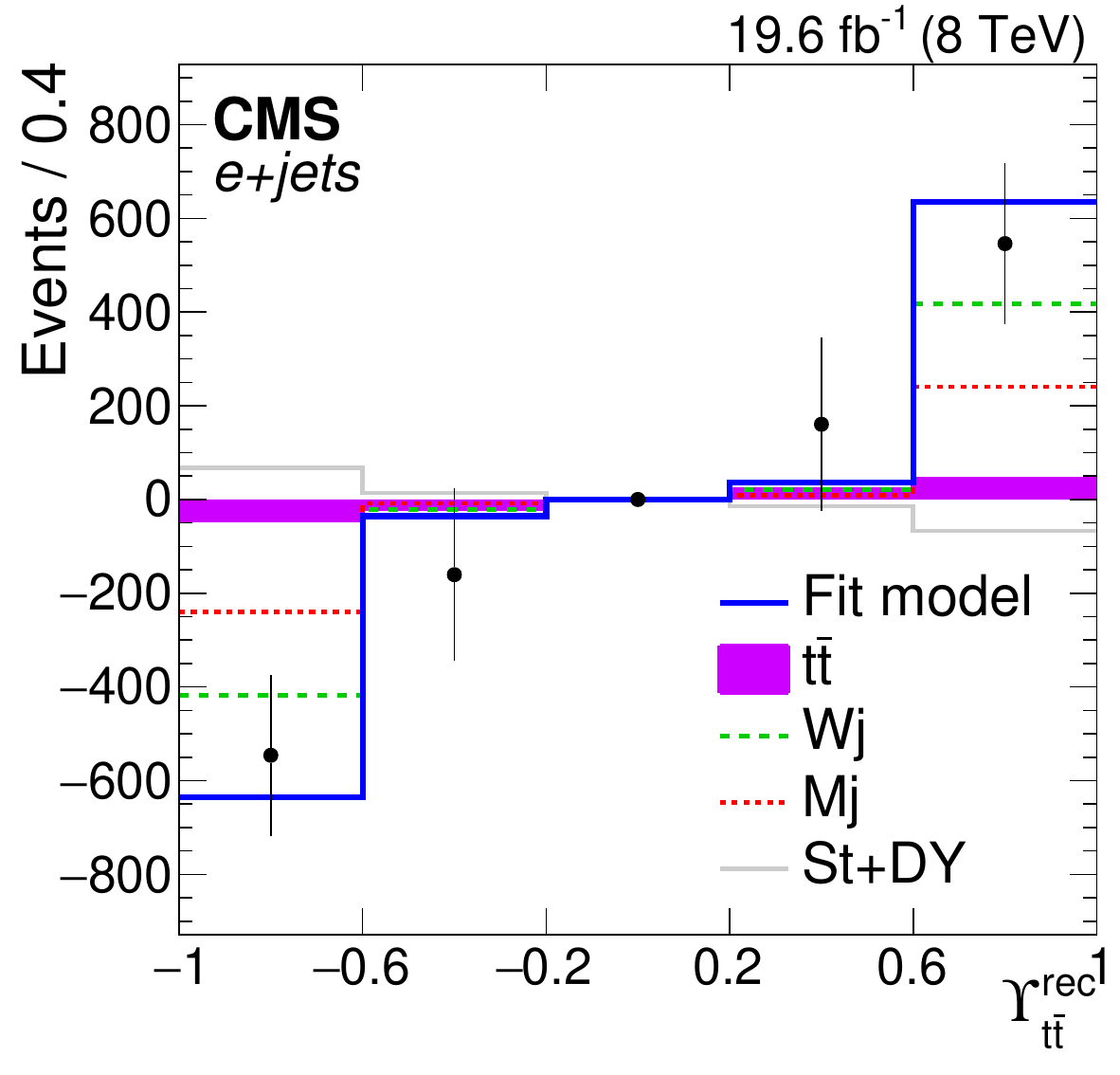}
  \quad
  \includegraphics[width=\projwidth]{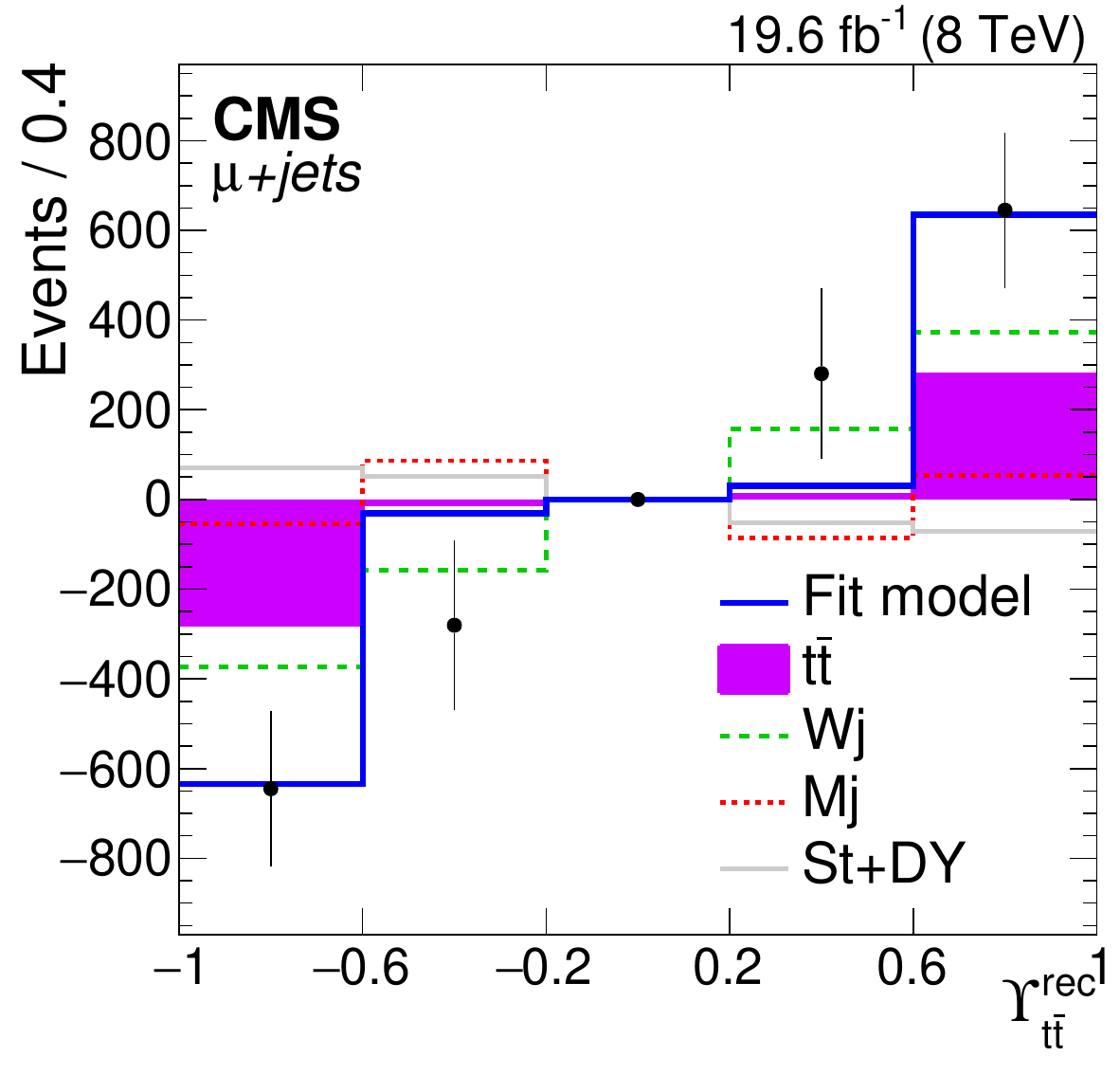}\\
  \caption{\label{results}
    Sample composition is measured using the discriminant $\triD$
    distribution (top), in a model with contributions from $\ttbar$,
    \Wj, \MJ, and \ST + \DY.
    With the sample composition subsequently fixed, the amplitude of
    the antisymmetric $\ttbar$ contribution is measured in the $\XLreco$
    distribution, shown decomposed into symmetric (middle) and
    antisymmetric (bottom) components.
    The thick line shows the antisymmetric component of the fit model.
    The measurements are performed independently on the (left) \ejets\
    and (right) \mujets\ samples.
  }
\end{figure*}

The measured sample composition is presented in
\Tab{sample_composition}.
\begin{table*}[htb]
  \centering
  \topcaption{\label{sample_composition}
Results from the fit of the sample composition, in thousands of
events, for the \ejets\ and \mujets\ channels.
The statistical uncertainty in the last digits is indicated in parentheses.
The results of the simultaneous fit in both channels are included only
for comparison and are not used in the measurement of $A_c^{y}$.
}
  \begin{scotch}{r..l.llr}
    & \multicolumn{7}{c}{Thousands of events}\\\cline{2-8}\\[-2.2ex]
    & \multicolumn{1}{c}{\ttbar}
    & \multicolumn{1}{c}{\Wj}
    & \multicolumn{1}{c}{\MJ}
    & \multicolumn{1}{c}{\ST}
    & \multicolumn{1}{c}{\DY}
    & \multicolumn{1}{c}{Total}
    & \multicolumn{1}{r}{Observed} \\
    \hline
    \Pe only
     & 207.1(8) &  49.1(9) &  50.5(1.1) & 14.0 & 5.4 &   326.2(1.6) & 326.185 \\
    \PGm only
     & 242.5(8) &  58.9(6) &  18.7(5) & 16.5 & 4.3 &   340.8(1.1) & 340.911 \\[2ex]
    Simultaneous fit
    \Pe & 207.1(5) &  49.5(4) &  50.2(6) & 14.0 & 5.4 & 326.2(9) & 326.185 \\
    \PGm & 242.6(6) &  58.8(5) &  18.7(5) & 16.5 & 4.3 & 340.9(9) & 340.911 \\ \end{scotch}
\end{table*}
Figure \ref{results} shows the data from each channel projected along
$\XLreco$ and $\triD$, overlaid with the results of the fitted model.

\begin{figure*}[htbp]
  \centering
  \includegraphics[width=0.55\textwidth]{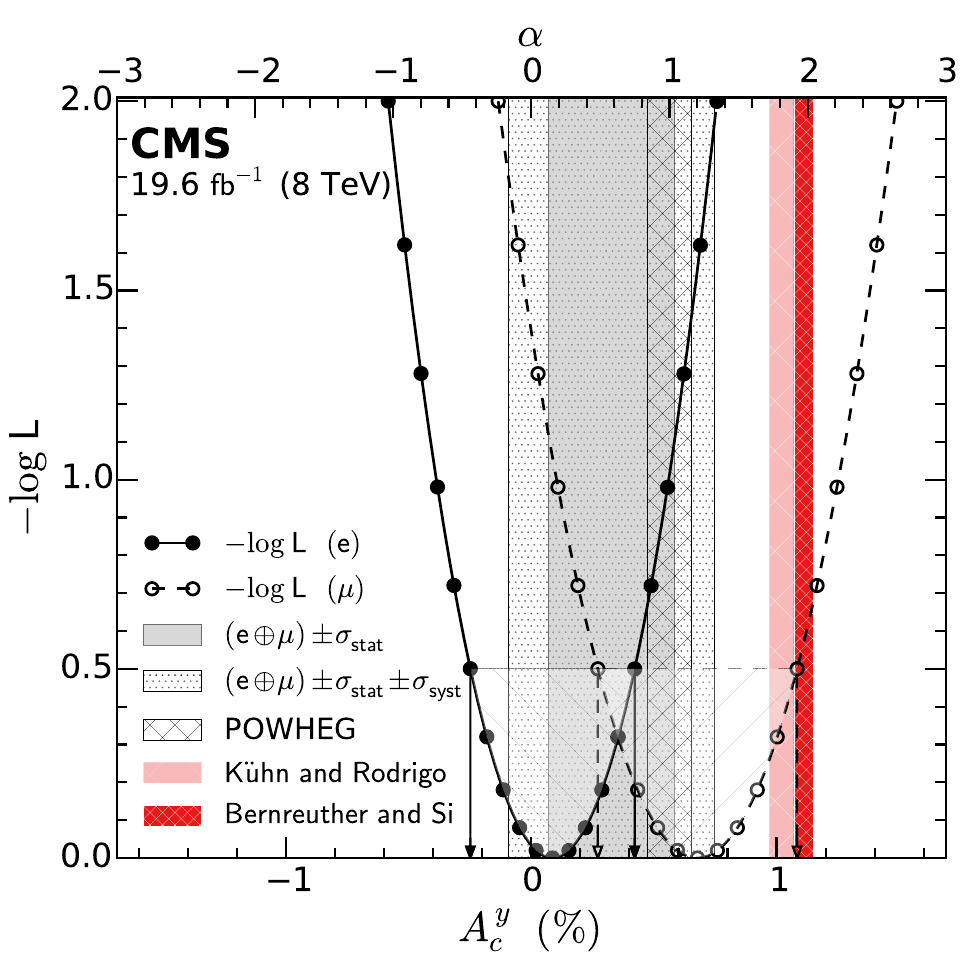}\\
  \includegraphics[width=0.48\textwidth]{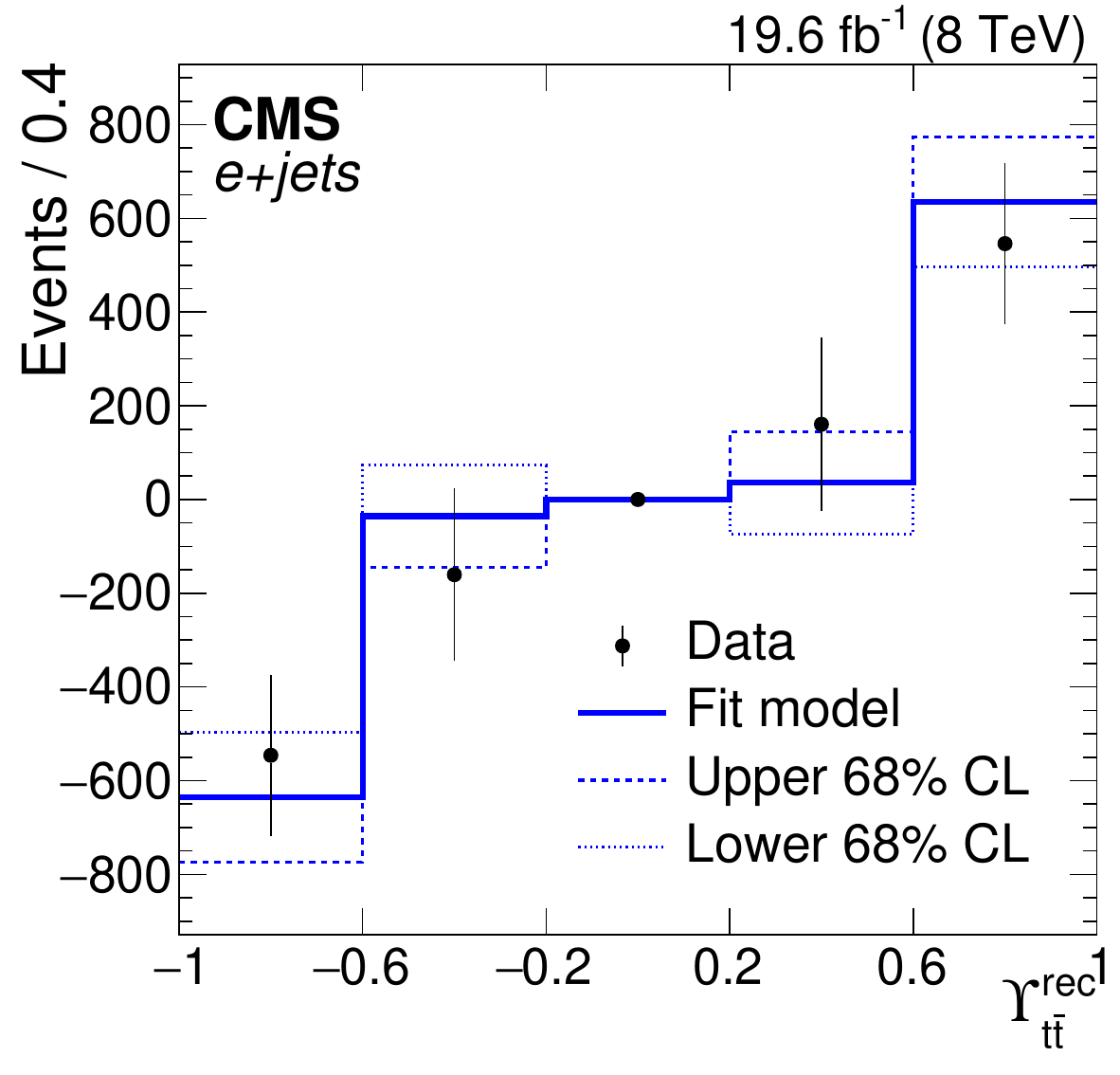}
  \quad
  \includegraphics[width=0.48\textwidth]{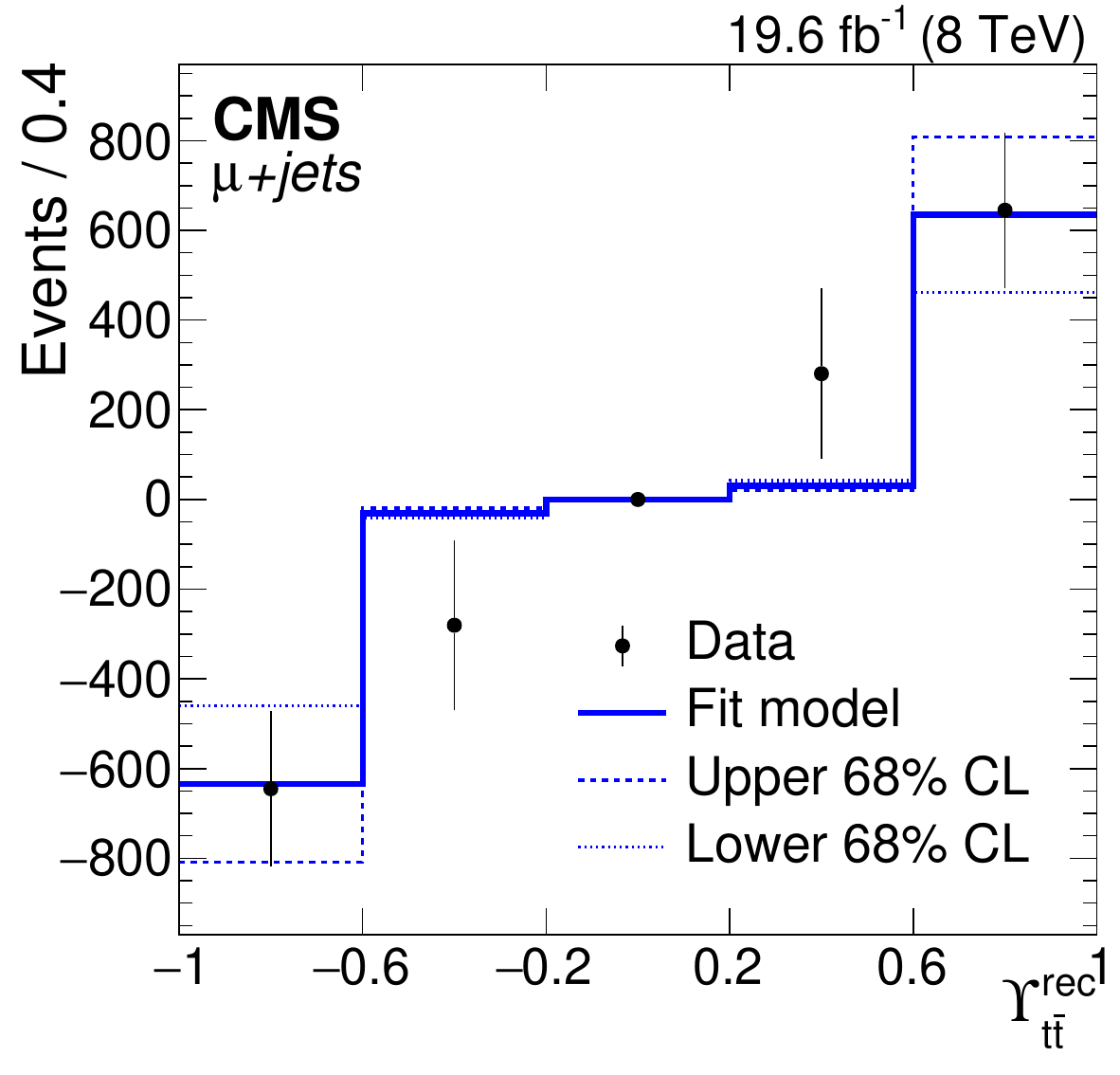}
  \caption{\label{uncertainties_plot} At top, the negative logarithm
    of the likelihood is shown as a function of $\alpha$ (upper axis)
    and $A_c^y$ (lower axis), for \ejets\ (closed circles) and \mujets\ (open circles) measurements.
    The statistical uncertainty in each is given by the intersections
    of the parabolas with $-\log L=0.5$, which are marked by arrows.
    The 68\% confidence interval of the combined $A_c^y$ measurement
    is compared with those of the SM predictions by \POWHEG,
    \KnR~\cite{Kuhn:2011ri}, and \BnS~\cite{PhysRevD.86.034026}. At bottom, the
    antisymmetric component of the $\XLreco$ distributions in data and
    the model are shown for (left) \ejets\ and (right) \mujets, for
    the central value (solid), and for the upper (dashed) and lower (dotted)
    limits of the 68\% statistical confidence intervals. }
\end{figure*}

Curves of the negative logarithm of the likelihood for both channels
are shown in \Fig{uncertainties_plot}, along with the combined $68\%$ confidence
interval for $A_c^y$.
The predictions of \POWHEG, \KnR~\cite{Kuhn:2011ri}, and
\BnS~\cite{PhysRevD.86.034026} are also plotted.
Subfigures of \Fig{uncertainties_plot} show the range of the
antisymmetric components covered by the models at ${\pm}1$ standard deviation of
the statistical uncertainty.
The combined charge asymmetry using both channels is
$A_c^y=[0.33\pm0.26\stat\pm0.33\syst]\%$,
which is tabulated with the predictions in \Tab{list_results}.
The combined uncertainty is 0.42\%.
\begin{table}[htb]
  \centering
  \topcaption{\label{list_results}
Comparison of charge asymmetry measurements and predictions.
}
  \begin{scotch}{cx}
    Source & \multicolumn{1}{c}{$A_c^{y} (\%)$} \\
    \hline
    \ejets                        &  0.09,0.34\stat\\
    \mujets                        &  0.68,0.41\stat\\
    Combined                            &  0.33,0.26\stat\pm0.33\syst \\[1ex]
    \POWHEG CT10                   &  0.56,0.09\\
    \MCATNLO                            &  0.53,0.09 \\
    \KnR~\cite{Kuhn:2011ri}        & 1.02,0.05 \\
    \BnS~\cite{PhysRevD.86.034026} & 1.11,0.04 \\
  \end{scotch}
\end{table}

\begin{figure}[htb]
 \centering
  \includegraphics[width=\cmsFigWidth]{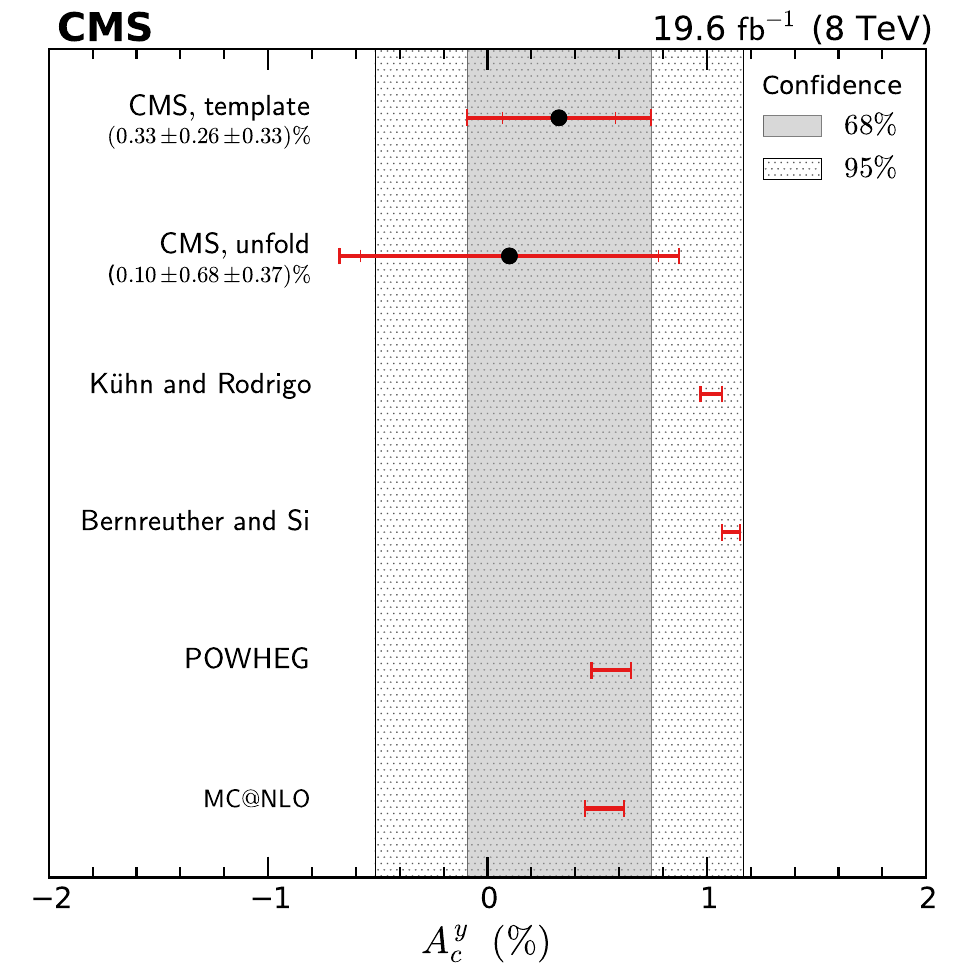}
  \caption{\label{result_plot} Comparison of results from this
    analysis (template) with those of the CMS 8\TeV unfolding
    analysis~\cite{unfold8TeV}, and SM predictions from theoretical
    calculations of \KnR~\cite{Kuhn:2011ri},
    \BnS~\cite{PhysRevD.86.034026}, \POWHEG, and \MCATNLO.  The shaded
    bands correspond to 68\% and 95\% confidence intervals of the
    current measurement. The inner bars on the CMS measurements
    indicate the statistical uncertainty,
    the outer bars the statistical and systematic uncertainty added in quadrature.
}
\end{figure}

The measured \ttbar production charge asymmetry $A_c^{y}$ is
compatible with another CMS $\sqrt{s}=$ 8\TeV
measurement~\cite{unfold8TeV}, which uses an unfolding technique on
the same data, and with the most recent Monte Carlo predictions and
theoretical calculations.
The template method incorporates more information from the model than
used in comparable unfolding
techniques~\cite{Chatrchyan201228,Chatrchyan2012129,ATLASTTBAR,Aad:2013cea,unfold8TeV} by
using the distribution of the antisymmetric component of the probability
density.
This extra information carries the benefit of reduced statistical
uncertainty, at the expense of greater model dependence, reflected in
the systematic uncertainty.
The contributions to the uncertainty from statistical and
systematic sources are comparable in size.
Since the systematic uncertainty is dominated by the statistical
uncertainty in the templates, it can be reduced in future analyses
through increased numbers of events in the simulation and in the sidebands in
the data.
The uncertainty in the \POWHEG prediction arises from systematic
uncertainties in the PDFs, the renormalization and factorization
scales, and the strong coupling constant.
A graphical comparison of the results and predictions is shown in
\Fig{result_plot}.

\section{Summary}
\label{summary}

The forward-central \ttbar charge asymmetry in proton-proton
collisions at $8\TeV$ center-of-mass energy has been measured using
lepton+jets events from data corresponding to an integrated
luminosity of 19.6$\fbinv$.
Novel techniques in top quark reconstruction and background
discrimination have been employed, which are likely to be of interest in
future analyses.
The measurement utilizes a template technique based on a
parametrization of the SM.
The result,
$A_c^y = [0.33\pm0.26\stat\pm0.33\syst]\%$,
is the most precise to date.
It is consistent with SM predictions, but does not rule out the
alternative models considered.

\begin{acknowledgments}
\hyphenation{Bundes-ministerium Forschungs-gemeinschaft Forschungs-zentren} We congratulate our colleagues in the CERN accelerator departments for the excellent performance of the LHC and thank the technical and administrative staffs at CERN and at other CMS institutes for their contributions to the success of the CMS effort. In addition, we gratefully acknowledge the computing centers and personnel of the Worldwide LHC Computing Grid for delivering so effectively the computing infrastructure essential to our analyses. Finally, we acknowledge the enduring support for the construction and operation of the LHC and the CMS detector provided by the following funding agencies: the Austrian Federal Ministry of Science, Research and Economy and the Austrian Science Fund; the Belgian Fonds de la Recherche Scientifique, and Fonds voor Wetenschappelijk Onderzoek; the Brazilian Funding Agencies (CNPq, CAPES, FAPERJ, and FAPESP); the Bulgarian Ministry of Education and Science; CERN; the Chinese Academy of Sciences, Ministry of Science and Technology, and National Natural Science Foundation of China; the Colombian Funding Agency (COLCIENCIAS); the Croatian Ministry of Science, Education and Sport, and the Croatian Science Foundation; the Research Promotion Foundation, Cyprus; the Ministry of Education and Research, Estonian Research Council via IUT23-4 and IUT23-6 and European Regional Development Fund, Estonia; the Academy of Finland, Finnish Ministry of Education and Culture, and Helsinki Institute of Physics; the Institut National de Physique Nucl\'eaire et de Physique des Particules~/~CNRS, and Commissariat \`a l'\'Energie Atomique et aux \'Energies Alternatives~/~CEA, France; the Bundesministerium f\"ur Bildung und Forschung, Deutsche Forschungsgemeinschaft, and Helmholtz-Gemeinschaft Deutscher Forschungszentren, Germany; the General Secretariat for Research and Technology, Greece; the National Scientific Research Foundation, and National Innovation Office, Hungary; the Department of Atomic Energy and the Department of Science and Technology, India; the Institute for Studies in Theoretical Physics and Mathematics, Iran; the Science Foundation, Ireland; the Istituto Nazionale di Fisica Nucleare, Italy; the Ministry of Science, ICT and Future Planning, and National Research Foundation (NRF), Republic of Korea; the Lithuanian Academy of Sciences; the Ministry of Education, and University of Malaya (Malaysia); the Mexican Funding Agencies (CINVESTAV, CONACYT, SEP, and UASLP-FAI); the Ministry of Business, Innovation and Employment, New Zealand; the Pakistan Atomic Energy Commission; the Ministry of Science and Higher Education and the National Science Centre, Poland; the Funda\c{c}\~ao para a Ci\^encia e a Tecnologia, Portugal; JINR, Dubna; the Ministry of Education and Science of the Russian Federation, the Federal Agency of Atomic Energy of the Russian Federation, Russian Academy of Sciences, and the Russian Foundation for Basic Research; the Ministry of Education, Science and Technological Development of Serbia; the Secretar\'{\i}a de Estado de Investigaci\'on, Desarrollo e Innovaci\'on and Programa Consolider-Ingenio 2010, Spain; the Swiss Funding Agencies (ETH Board, ETH Zurich, PSI, SNF, UniZH, Canton Zurich, and SER); the Ministry of Science and Technology, Taipei; the Thailand Center of Excellence in Physics, the Institute for the Promotion of Teaching Science and Technology of Thailand, Special Task Force for Activating Research and the National Science and Technology Development Agency of Thailand; the Scientific and Technical Research Council of Turkey, and Turkish Atomic Energy Authority; the National Academy of Sciences of Ukraine, and State Fund for Fundamental Researches, Ukraine; the Science and Technology Facilities Council, UK; the US Department of Energy, and the US National Science Foundation.

Individuals have received support from the Marie-Curie program and the European Research Council and EPLANET (European Union); the Leventis Foundation; the A. P. Sloan Foundation; the Alexander von Humboldt Foundation; the Belgian Federal Science Policy Office; the Fonds pour la Formation \`a la Recherche dans l'Industrie et dans l'Agriculture (FRIA-Belgium); the Agentschap voor Innovatie door Wetenschap en Technologie (IWT-Belgium); the Ministry of Education, Youth and Sports (MEYS) of the Czech Republic; the Council of Science and Industrial Research, India; the HOMING PLUS program of the Foundation for Polish Science, cofinanced from European Union, Regional Development Fund; the OPUS program of the National Science Center (Poland); the Compagnia di San Paolo (Torino); the Consorzio per la Fisica (Trieste); MIUR project 20108T4XTM (Italy); the Thalis and Aristeia programs cofinanced by EU-ESF and the Greek NSRF; the National Priorities Research Program by Qatar National Research Fund; the Rachadapisek Sompot Fund for Postdoctoral Fellowship, Chulalongkorn University (Thailand); and the Welch Foundation, contract C-1845.
\end{acknowledgments}
\bibliography{auto_generated}

\providecommand{\href}[2]{#2}\begingroup\raggedright\begin{thebibliography}{10}%
\makeatletter
\providecommand{\hrefCMSnoop }[0]{\@secondoftwo}%
\makeatother
\providecommand{\doi}{\texttt{doi:}\begingroup \urlstyle{tt}\Url}

\bibitem{Agashe:2014kda}
\hrefCMSnoop {}{{Particle Data Group} Collaboration, ``{Review of Particle
  Physics}'',} \textit{ Chin. Phys. C} \textbf{ 38} (2014) 090001,
\href{http://dx.doi.org/10.1088/1674-1137/38/9/090001}{\doi{10.1088/1674-1137/38/9/090001}}.

\bibitem{Frampton:1987dn}
\hrefCMSnoop {}{P.~H. Frampton and S.~L. Glashow, ``{Chiral Color: An
  Alternative to the Standard Model}'',} \textit{ Phys. Lett. B} \textbf{ 190}
  (1987) 157,
\href{http://dx.doi.org/10.1016/0370-2693(87)90859-8}{\doi{10.1016/0370-2693(87)90859-8}}.

\bibitem{Budny:1973dj}
\hrefCMSnoop {}{R.~Budny, ``{Effects of neutral weak currents in
  annihilation}'',} \textit{ Phys. Lett. B} \textbf{ 45} (1973) 340,
\href{http://dx.doi.org/10.1016/0370-2693(73)90050-6}{\doi{10.1016/0370-2693(73)90050-6}}.

\bibitem{Abazov:2011rq}
\hrefCMSnoop {}{{D0} Collaboration, ``{Forward-backward asymmetry in top
  quark-antiquark production}'',} \textit{ Phys. Rev. D} \textbf{ 84} (2011)
  112005,
  \href{http://dx.doi.org/10.1103/PhysRevD.84.112005}{\doi{10.1103/PhysRevD.84.112005}},
\href{http://www.arXiv.org/abs/1107.4995}{\texttt{arXiv:1107.4995}}.

\bibitem{Aaltonen:2011kc}
\hrefCMSnoop {}{{CDF} Collaboration, ``{Evidence for a Mass Dependent
  Forward-Backward Asymmetry in Top Quark Pair Production}'',} \textit{ Phys.
  Rev. D} \textbf{ 83} (2011) 112003,
  \href{http://dx.doi.org/10.1103/PhysRevD.83.112003}{\doi{10.1103/PhysRevD.83.112003}},
\href{http://www.arXiv.org/abs/1101.0034}{\texttt{arXiv:1101.0034}}.

\bibitem{PhysRevD.87.092002}
\hrefCMSnoop {}{{CDF} Collaboration, ``Measurement of the top quark
  forward-backward production asymmetry and its dependence on event kinematic
  properties'',} \textit{ Phys. Rev. D} \textbf{ 87} (2013) 092002,
  \href{http://dx.doi.org/10.1103/PhysRevD.87.092002}{\doi{10.1103/PhysRevD.87.092002}},
  \href{http://www.arXiv.org/abs/1211.1003}{\texttt{arXiv:1211.1003}}.

\bibitem{Abazov:2014cca}
\hrefCMSnoop {}{{D0} Collaboration, ``{Measurement of the forward-backward
  asymmetry in top quark-antiquark production in $p\bar{p}$ collisions using
  the lepton $+$ jets channel}'',} \textit{ Phys. Rev. D} \textbf{ 90} (2014)
  072011,
  \href{http://dx.doi.org/10.1103/PhysRevD.90.072011}{\doi{10.1103/PhysRevD.90.072011}},
\href{http://www.arXiv.org/abs/1405.0421}{\texttt{arXiv:1405.0421}}.

\bibitem{Kuhn:2011ri}
\hrefCMSnoop {}{J.~H. K{\"u}hn and G.~Rodrigo, ``{Charge asymmetries of top
  quarks at hadron colliders revisited}'',} \textit{ JHEP} \textbf{ 01} (2012)
  063,
  \href{http://dx.doi.org/10.1007/JHEP01(2012)063}{\doi{10.1007/JHEP01(2012)063}},
\href{http://www.arXiv.org/abs/1109.6830}{\texttt{arXiv:1109.6830}}.

\bibitem{PhysRevD.86.034026}
\hrefCMSnoop {}{W.~Bernreuther and Z.-G. Si, ``Top quark and leptonic charge
  asymmetries for the {Tevatron} and {LHC}'',} \textit{ Phys. Rev. D} \textbf{
  86} (2012) 034026,
  \href{http://dx.doi.org/10.1103/PhysRevD.86.034026}{\doi{10.1103/PhysRevD.86.034026}},
  \href{http://www.arXiv.org/abs/1205.6580}{\texttt{arXiv:1205.6580}}.

\bibitem{Chatrchyan:2013qha}
\hrefCMSnoop {}{{CMS Collaboration}, ``{Search for narrow resonances using the
  dijet mass spectrum in $pp$ collisions at $\sqrt{s} = 8$~TeV}'',} \textit{
  Phys. Rev. D} \textbf{ 87} (2013) 114015,
  \href{http://dx.doi.org/10.1103/PhysRevD.87.114015}{\doi{10.1103/PhysRevD.87.114015}},
\href{http://www.arXiv.org/abs/1302.4794}{\texttt{arXiv:1302.4794}}.

\bibitem{Aad:2014aqa}
\hrefCMSnoop {}{{ATLAS Collaboration}, ``{Search for new phenomena in the dijet
  mass distribution using pp collision data at $\sqrt{s}$ = 8 TeV with the
  ATLAS detector}'',} \textit{ Phys. Rev. D} \textbf{ 91} (2015) 052007,
  \href{http://dx.doi.org/10.1103/PhysRevD.91.052007}{\doi{10.1103/PhysRevD.91.052007}},
\href{http://www.arXiv.org/abs/1407.1376}{\texttt{arXiv:1407.1376}}.

\bibitem{ATLAS:2012pu}
\hrefCMSnoop {}{{ATLAS Collaboration}, ``{ATLAS search for new phenomena in
  dijet mass and angular distributions using pp collisions at $\sqrt{s}$ = 7
  TeV}'',} \textit{ JHEP} \textbf{ 01} (2013) 029,
  \href{http://dx.doi.org/10.1007/JHEP01(2013)029}{\doi{10.1007/JHEP01(2013)029}},
\href{http://www.arXiv.org/abs/1210.1718}{\texttt{arXiv:1210.1718}}.

\bibitem{benchmarks}
A.~Carmona\hrefCMSnoop {}{ {et~al.}, ``{From Tevatron's top and lepton-based
  asymmetries to the LHC}'',} \textit{ JHEP} \textbf{ 07} (2014) 005,
  \href{http://dx.doi.org/10.1007/JHEP07(2014)005}{\doi{10.1007/JHEP07(2014)005}},
\href{http://www.arXiv.org/abs/1401.2443}{\texttt{arXiv:1401.2443}}.

\bibitem{Atre:2013mja}
\hrefCMSnoop {}{A.~Atre, R.~S. Chivukula, P.~Ittisamai, and E.~H. Simmons,
  ``{Distinguishing color-octet and color-singlet resonances at the Large
  Hadron Collider}'',} \textit{ Phys. Rev. D} \textbf{ 88} (2013) 055021,
  \href{http://dx.doi.org/10.1103/PhysRevD.88.055021}{\doi{10.1103/PhysRevD.88.055021}},
\href{http://www.arXiv.org/abs/1306.4715}{\texttt{arXiv:1306.4715}}.

\bibitem{Chatrchyan201228}
\hrefCMSnoop {}{{CMS Collaboration}, ``{Measurement of the charge asymmetry in
  top-quark pair production in proton-proton collisions at $\sqrt{s}$ = 7
  TeV}'',} \textit{ Phys. Lett. B} \textbf{ 709} (2012) 28,
  \href{http://dx.doi.org/10.1016/j.physletb.2012.01.078}{\doi{10.1016/j.physletb.2012.01.078}},
\href{http://www.arXiv.org/abs/1112.5100}{\texttt{arXiv:1112.5100}}.

\bibitem{Chatrchyan2012129}
\hrefCMSnoop {}{{CMS Collaboration}, ``{Inclusive and differential measurements
  of the $\mathrm{t}\overline{\mathrm{t}}$ charge asymmetry in proton-proton
  collisions at $\sqrt{s}$ = 7 TeV}'',} \textit{ Phys. Lett. B} \textbf{ 717}
  (2012) 129,
  \href{http://dx.doi.org/10.1016/j.physletb.2012.09.028}{\doi{10.1016/j.physletb.2012.09.028}},
\href{http://www.arXiv.org/abs/1207.0065}{\texttt{arXiv:1207.0065}}.

\bibitem{ATLASTTBAR}
\hrefCMSnoop {}{{ATLAS Collaboration}, ``{Measurement of the charge asymmetry
  in top quark pair production in pp collisions at $\sqrt{s}$ = 7 TeV using the
  ATLAS detector}'',} \textit{ Eur. Phys. J. C} \textbf{ 72} (2012) 2039,
  \href{http://dx.doi.org/10.1140/epjc/s10052-012-2039-5}{\doi{10.1140/epjc/s10052-012-2039-5}},
\href{http://www.arXiv.org/abs/1203.4211}{\texttt{arXiv:1203.4211}}.

\bibitem{Aad:2013cea}
\hrefCMSnoop {}{{ATLAS Collaboration}, ``{Measurement of the top quark pair
  production charge asymmetry in proton-proton collisions at $\sqrt{s}$ = 7 TeV
  using the ATLAS detector}'',} \textit{ JHEP} \textbf{ 02} (2014) 107,
  \href{http://dx.doi.org/10.1007/JHEP02(2014)107}{\doi{10.1007/JHEP02(2014)107}},
\href{http://www.arXiv.org/abs/1311.6724}{\texttt{arXiv:1311.6724}}.

\bibitem{unfold8TeV}
\hrefCMSnoop {}{{CMS Collaboration}, ``{Inclusive and differential measurements
  of the $\mathrm{t}\overline{\mathrm{t}}$ charge asymmetry in pp collisions at
  $\sqrt{s}$ = 8 TeV}'',} (2015).
  \href{http://www.arXiv.org/abs/1507.03119}{\texttt{arXiv:1507.03119}}.
Submitted to \textit{Phys. Lett. B}.

\bibitem{Frixione:2007nw}
\hrefCMSnoop {}{S.~Frixione, P.~Nason, and G.~Ridolfi, ``{A positive-weight
  next-to-leading-order Monte Carlo for heavy flavour hadroproduction}'',}
  \textit{ JHEP} \textbf{ 09} (2007) 126,
  \href{http://dx.doi.org/10.1088/1126-6708/2007/09/126}{\doi{10.1088/1126-6708/2007/09/126}},
\href{http://www.arXiv.org/abs/0707.3088}{\texttt{arXiv:0707.3088}}.

\bibitem{Lai:2010vv}
H.-L. Lai\hrefCMSnoop {}{ {et~al.}, ``New parton distributions for collider
  physics'',} \textit{ Phys. Rev. D} \textbf{ 82} (2010) 074024,
  \href{http://dx.doi.org/10.1103/PhysRevD.82.074024}{\doi{10.1103/PhysRevD.82.074024}},
\href{http://www.arXiv.org/abs/1007.2241}{\texttt{arXiv:1007.2241}}.

\bibitem{CMS-PAS-PFT-09-001}
\href {http://cdsweb.cern.ch/record/1194487}{{CMS Collaboration},
  ``Particle--flow event reconstruction in {CMS} and performance for jets,
  taus, and {\MET}'',} CMS Physics Analysis Summary CMS-PAS-PFT-09-001, 2009.

\bibitem{CMS-PAS-PFT-10-001}
\href {http://cdsweb.cern.ch/record/1247373}{{CMS Collaboration},
  ``Commissioning of the particle-flow event reconstruction with the first
  {LHC} collisions recorded in the {CMS} detector'',} CMS Physics Analysis
  Summary CMS-PAS-PFT-10-001, 2010.

\bibitem{Khachatryan:2015hwa}
\hrefCMSnoop {}{{CMS Collaboration}, ``{Performance of electron reconstruction
  and selection with the CMS detector in proton-proton collisions at $\sqrt{s}$
  = 8 TeV}'',} \textit{ JINST} \textbf{ 10} (2015) P06005,
  \href{http://dx.doi.org/10.1088/1748-0221/10/06/P06005}{\doi{10.1088/1748-0221/10/06/P06005}},
\href{http://www.arXiv.org/abs/1502.02701}{\texttt{arXiv:1502.02701}}.

\bibitem{Cacciari:2008gp}
\hrefCMSnoop {}{M.~Cacciari, G.~P. Salam, and G.~Soyez, ``The anti-$k_t$ jet
  clustering algorithm'',} \textit{ JHEP} \textbf{ 04} (2008) 063,
  \href{http://dx.doi.org/10.1088/1126-6708/2008/04/063}{\doi{10.1088/1126-6708/2008/04/063}},
  \href{http://www.arXiv.org/abs/0802.1189}{\texttt{arXiv:0802.1189}}.

\bibitem{Chatrchyan:2011ds}
\hrefCMSnoop {}{{CMS Collaboration}, ``{Determination of jet energy calibration
  and transverse momentum resolution in CMS}'',} \textit{ JINST} \textbf{ 6}
  (2011) P11002,
  \href{http://dx.doi.org/10.1088/1748-0221/6/11/P11002}{\doi{10.1088/1748-0221/6/11/P11002}},
\href{http://www.arXiv.org/abs/1107.4277}{\texttt{arXiv:1107.4277}}.

\bibitem{1748-0221-8-04-P04013}
\hrefCMSnoop {}{{CMS Collaboration}, ``{Identification of b-quark jets with the
  CMS experiment}'',} \textit{ JINST} \textbf{ 8} (2013) P04013,
  \href{http://dx.doi.org/10.1088/1748-0221/8/04/P04013}{\doi{10.1088/1748-0221/8/04/P04013}},
\href{http://www.arXiv.org/abs/1211.4462}{\texttt{arXiv:1211.4462}}.

\bibitem{Chatrchyan:2008zzk}
\hrefCMSnoop {}{{CMS Collaboration}, ``The {CMS} experiment at the {CERN}
  {LHC}'',} \textit{ JINST} \textbf{ 3} (2008) S08004,
\href{http://dx.doi.org/10.1088/1748-0221/3/08/S08004}{\doi{10.1088/1748-0221/3/08/S08004}}.

\bibitem{Leonard:2014qda}
\href {http://pos.sissa.it/archive/conferences/213/348/TIPP2014_348.pdf}{J.~L.
  Leonard, ``{Luminosity measurement at CMS}'',} in \textit{ {Proceedings, 3rd
  International Conference on Technology and Instrumentation in Particle
  Physics (TIPP 2014)}}, p.~348.
\newblock {SISSA}, 2014.
\newblock
{PoS(TIPP2014)348}.

\bibitem{Chatrchyan:2012xi}
\hrefCMSnoop {}{{CMS Collaboration}, ``{Performance of CMS muon reconstruction
  in pp collision events at $\sqrt{s}$ = 7 TeV}'',} \textit{ JINST} \textbf{ 7}
  (2012) P10002,
  \href{http://dx.doi.org/10.1088/1748-0221/7/10/P10002}{\doi{10.1088/1748-0221/7/10/P10002}},
\href{http://www.arXiv.org/abs/1206.4071}{\texttt{arXiv:1206.4071}}.

\bibitem{Agostinelli:2002hh}
\hrefCMSnoop {}{{GEANT4} Collaboration, ``{GEANT4}---simulation toolkit'',}
  \textit{ Nucl. Instrum. Meth. A} \textbf{ 506} (2003) 250,
\href{http://dx.doi.org/10.1016/S0168-9002(03)01368-8}{\doi{10.1016/S0168-9002(03)01368-8}}.

\bibitem{Sjostrand:2006za}
\hrefCMSnoop {}{T.~Sj{\"o}strand, S.~Mrenna, and P.~Z. Skands, ``{PYTHIA} 6.4
  physics and manual'',} \textit{ JHEP} \textbf{ 05} (2006) 026,
  \href{http://dx.doi.org/10.1088/1126-6708/2006/05/026}{\doi{10.1088/1126-6708/2006/05/026}},
\href{http://www.arXiv.org/abs/hep-ph/0603175}{\texttt{arXiv:hep-ph/0603175}}.

\bibitem{Sjostrand:2007gs}
\hrefCMSnoop {}{T.~Sj{\"o}strand, S.~Mrenna, and P.~Z. Skands, ``{A Brief
  Introduction to PYTHIA 8.1}'',} \textit{ Comput. Phys. Commun.} \textbf{ 178}
  (2008) 852,
  \href{http://dx.doi.org/10.1016/j.cpc.2008.01.036}{\doi{10.1016/j.cpc.2008.01.036}},
\href{http://www.arXiv.org/abs/0710.3820}{\texttt{arXiv:0710.3820}}.

\bibitem{Corke:2010zj}
\hrefCMSnoop {}{R.~Corke and T.~Sj{\"o}strand, ``{Improved Parton Showers at
  Large Transverse Momenta}'',} \textit{ Eur. Phys. J. C} \textbf{ 69} (2010)
  1,
  \href{http://dx.doi.org/10.1140/epjc/s10052-010-1409-0}{\doi{10.1140/epjc/s10052-010-1409-0}},
\href{http://www.arXiv.org/abs/1003.2384}{\texttt{arXiv:1003.2384}}.

\bibitem{Alwall:2011uj}
J.~Alwall\hrefCMSnoop {}{ {et~al.}, ``{MadGraph 5}: going beyond'',} \textit{
  JHEP} \textbf{ 06} (2011) 128,
  \href{http://dx.doi.org/10.1007/JHEP06(2011)128}{\doi{10.1007/JHEP06(2011)128}},
\href{http://www.arXiv.org/abs/1106.0522}{\texttt{arXiv:1106.0522}}.

\bibitem{Pumplin:2005rh}
J.~Pumplin\hrefCMSnoop {}{ {et~al.}, ``{Parton distributions and the strong
  coupling: CTEQ6AB PDFs}'',} \textit{ JHEP} \textbf{ 02} (2006) 032,
  \href{http://dx.doi.org/10.1088/1126-6708/2006/02/032}{\doi{10.1088/1126-6708/2006/02/032}},
\href{http://www.arXiv.org/abs/hep-ph/0512167}{\texttt{arXiv:hep-ph/0512167}}.

\bibitem{Alioli:2009je}
\hrefCMSnoop {}{S.~Alioli, P.~Nason, C.~Oleari, and E.~Re, ``{NLO single-top
  production matched with shower in POWHEG: $s$- and $t$-channel
  contributions}'',} \textit{ JHEP} \textbf{ 09} (2009) 111,
  \href{http://dx.doi.org/10.1088/1126-6708/2009/09/111}{\doi{10.1088/1126-6708/2009/09/111}},
  \href{http://www.arXiv.org/abs/0907.4076}{\texttt{arXiv:0907.4076}}.
[Erratum: \DOI{10.1007/JHEP02(2010)011}].

\bibitem{Re:2010bp}
\hrefCMSnoop {}{E.~Re, ``{Single-top $W$ $t$-channel production matched with
  parton showers using the POWHEG method}'',} \textit{ Eur. Phys. J. C}
  \textbf{ 71} (2011) 1547,
  \href{http://dx.doi.org/10.1140/epjc/s10052-011-1547-z}{\doi{10.1140/epjc/s10052-011-1547-z}},
\href{http://www.arXiv.org/abs/1009.2450}{\texttt{arXiv:1009.2450}}.

\bibitem{Frixione:2002ik}
\hrefCMSnoop {}{S.~Frixione and B.~R. Webber, ``{Matching NLO QCD computations
  and parton shower simulations}'',} \textit{ JHEP} \textbf{ 06} (2002) 029,
  \href{http://dx.doi.org/10.1088/1126-6708/2002/06/029}{\doi{10.1088/1126-6708/2002/06/029}},
\href{http://www.arXiv.org/abs/hep-ph/0204244}{\texttt{arXiv:hep-ph/0204244}}.

\bibitem{Giammanco:2014bza}
\hrefCMSnoop {}{A.~Giammanco, ``{The Fast Simulation of the CMS Experiment}'',}
  \textit{ J. Phys. Conf. Ser.} \textbf{ 513} (2014) 022012,
\href{http://dx.doi.org/10.1088/1742-6596/513/2/022012}{\doi{10.1088/1742-6596/513/2/022012}}.

\bibitem{Jung:2011zv}
\hrefCMSnoop {}{S.~Jung, A.~Pierce, and J.~D. Wells, ``{Top quark asymmetry
  from a non-Abelian horizontal symmetry}'',} \textit{ Phys. Rev. D} \textbf{
  83} (2011) 114039,
  \href{http://dx.doi.org/10.1103/PhysRevD.83.114039}{\doi{10.1103/PhysRevD.83.114039}},
\href{http://www.arXiv.org/abs/1103.4835}{\texttt{arXiv:1103.4835}}.

\bibitem{Betchart:2013nba}
\hrefCMSnoop {}{B.~A. Betchart, R.~Demina, and A.~Harel, ``Analytic solutions
  for neutrino momenta in decay of top quarks'',} \textit{ Nucl. Instrum. Meth.
  A} \textbf{ 736} (2014) 169,
  \href{http://dx.doi.org/10.1016/j.nima.2013.10.039}{\doi{10.1016/j.nima.2013.10.039}},
  \href{http://www.arXiv.org/abs/1305.1878}{\texttt{arXiv:1305.1878}}.

\bibitem{welch1939}
\hrefCMSnoop {}{B.~L. Welch, ``Note on discriminant functions'',} \textit{
  Biometrika} \textbf{ 31} (1939) 218,
  \href{http://dx.doi.org/10.2307/2334985}{\doi{10.2307/2334985}}.

\bibitem{Verkerke:2003ir}
\hrefCMSnoop {}{W.~Verkerke and D.~P. Kirkby, ``{The RooFit toolkit for data
  modeling}'',} in \textit{ Proceedings of the 13th Int. Conf. for Computing in
  High-Energy and Nuclear Phys.}
\newblock 2003.
\newblock
\href{http://www.arXiv.org/abs/physics/0306116}{\texttt{arXiv:physics/0306116}}.
\newblock

\bibitem{pdf4lhc}
M.~Botje\hrefCMSnoop {}{ {et~al.}, ``{The PDF4LHC Working Group Interim
  Recommendations}'',} (2011).
\href{http://www.arXiv.org/abs/1101.0538}{\texttt{arXiv:1101.0538}}.

\bibitem{Campbell:2010ff}
\hrefCMSnoop {}{J.~M. Campbell and R.~K. Ellis, ``{MCFM for the Tevatron and
  the LHC}'',} \textit{ Nucl. Phys. Proc. Suppl.} \textbf{ 205} (2010) 10,
  \href{http://dx.doi.org/10.1016/j.nuclphysbps.2010.08.011}{\doi{10.1016/j.nuclphysbps.2010.08.011}},
\href{http://www.arXiv.org/abs/1007.3492}{\texttt{arXiv:1007.3492}}.

\end{thebibliography}\endgroup

\cleardoublepage \appendix\section{The CMS Collaboration \label{app:collab}}\begin{sloppypar}\hyphenpenalty=5000\widowpenalty=500\clubpenalty=5000\textbf{Yerevan Physics Institute,  Yerevan,  Armenia}\\*[0pt]
V.~Khachatryan, A.M.~Sirunyan, A.~Tumasyan
\vskip\cmsinstskip
\textbf{Institut f\"{u}r Hochenergiephysik der OeAW,  Wien,  Austria}\\*[0pt]
W.~Adam, E.~Asilar, T.~Bergauer, J.~Brandstetter, E.~Brondolin, M.~Dragicevic, J.~Er\"{o}, M.~Flechl, M.~Friedl, R.~Fr\"{u}hwirth\cmsAuthorMark{1}, V.M.~Ghete, C.~Hartl, N.~H\"{o}rmann, J.~Hrubec, M.~Jeitler\cmsAuthorMark{1}, V.~Kn\"{u}nz, A.~K\"{o}nig, M.~Krammer\cmsAuthorMark{1}, I.~Kr\"{a}tschmer, D.~Liko, T.~Matsushita, I.~Mikulec, D.~Rabady\cmsAuthorMark{2}, B.~Rahbaran, H.~Rohringer, J.~Schieck\cmsAuthorMark{1}, R.~Sch\"{o}fbeck, J.~Strauss, W.~Treberer-Treberspurg, W.~Waltenberger, C.-E.~Wulz\cmsAuthorMark{1}
% ...
\vskip\cmsinstskip
% ...
1:~~Also at Vienna University of Technology, Vienna, Austria\\
2:~~Also at CERN, European Organization for Nuclear Research, Geneva, Switzerland\\
% ...

\end{sloppypar}
\end{document}